# Quantitative evaluation of regulatory policies for reducing deforestation using the bent-cable regression model

Megan C Evans[1], Grace Chiu[2], Philip Gibbons[3], Andrew K Macintosh[4]

Version 1.0
21st June 2019

## Abstract

Reducing and redressing the effects of deforestation is a complex public policy challenge, and evaluating the efficacy of such policy efforts is crucial for policy learning and adaptation. Deforestation in high-income nations can contribute substantially to global forest loss, despite the presence of strong institutions and high policy capacity. In Queensland, Australia, over 5 million hectares of native forest has been lost since 1988. Successive regulatory policies have aimed to reduce deforestation in Queensland, though debate exists over their effect given the influence of other drivers of forest loss. Using a hierarchical Bayesian statistical framework, we combine satellite imagery of forest loss with macroeconomic, land tenure, biophysical and climatic variables to collectively model deforestation for 50 local government areas (LGAs) across Queensland. We apply the spatially explicit bent-cable regression model to detect trend change that may signal a regulatory policy effect. We find that annual % growth in GDP was the only clear driver of LGA-specific deforestation after adjusting for other covariate effects. Our model shows strong evidence of spatial contagion in deforestation across Queensland, and this effect is influenced by the dominant land tenure type within each LGA. We find our model exhibits a "bend" mostly between 2000 and 2007, consistent with expectations, but the signal is not particularly strong due extreme variation in deforestation trends between and within LGAs. Our results demonstrate that the bent-cable model is a promising technique for detecting system changes in response to policy interventions, but future work should be conducted at a national scale to provide more data points, and incorporate more LGA-specific data to improve model goodness-of-fit.

* Corresponding author: megan.evans@uq.edu.au
[1] Centre for Biodiversity and Conservation Science and Centre for Policy Futures, The University of Queensland, Brisbane, Australia
[2] Virginia Institute of Marine Science, College of William & Mary, Virginia, United States of America
[3] Fenner School of Environment and Society, ANU College of Medicine, Biology and Environment, The Australian National University, Canberra, Australia
[4] ANU Centre for Climate Law and Policy, ANU College of Law, The Australian National University, Canberra, Australia

*Key words:* deforestation, regulation, environmental policy, bent-cable regression, Bayesian inference, hierarchical modeling, longitudinal data, spatial correlation, mixed effects model, nonlinear model,

**Highlights**

- Deforestation can be globally significant in high-income nations with strong governance
- Over 5 million hectares of forest has been cleared in Queensland, Australia, since 1988
- A spatially explicit bent-cable regression was used to model deforestation
- Strong evidence for spatial contagion in deforestation influenced by LGA land tenure
- Some evidence of state-wide policy effect due to extreme variation in LGA deforestation responses

**Introduction**

Effective control of deforestation is crucial to ensure sustainable provision of ecosystem services and the conservation of biodiversity [1–4]. Public policies such as protected areas [5–9], regulations and market intervention [10–15] and payments for ecosystem services [16–18] can contribute to forest protection. Not all policies have proved effective however, with some producing perverse outcomes [19,20] and many more policy interventions having not been evaluated for their efficacy [21–23]. An understanding of the efficacy of policy measures taken to control deforestation, and the varied institutional, social and political conditions in which they are adopted, implemented and enforced [24,25] is critical for policy learning and adaptation.

Disentangling the effects of policy interventions from the varied drivers of deforestation is a complex evaluation challenge. Broader macroeconomic trends and policy drivers ultimately influence local market conditions and institutional settings, which in turn affect the deforestation behavior of agents [26–28]. The availability of arable land and forest resources also influences how much deforestation occurs locally, and how much is displaced elsewhere [1,19,29]. How these variables interact and influence deforestation in a particular context cannot easily be generalized [27], hence there is a need to evaluate the efficacy of a broad range of instruments across multiple locations and policy settings [22].

Much of the published work on deforestation drivers and policy interventions has been in the tropics, perhaps due to the high rates of deforestation observed in this region [30,31], and the global significance of tropical forests for biodiversity, livelihoods, climate regulation and food production [32–34]. Population growth [35,36], roads and access to markets [2,37], agricultural commodity prices and currency exchange rates [38–40], presence and strength of institutions [9,41–44] and degree of policy enforcement [12,45] have all been found to influence tropical deforestation. However, less attention has been paid to deforestation drivers and policy responses in high income nations [46], which have contributed substantially to global forest loss [30,47] despite having comparatively strong governance [48,49]. Australia is one such example, where recent globally significant deforestation rates [31,50] have occurred against a backdrop of forest policy reform and associated political and social contestation [1,51–53]

Forest loss in Australia at the turn of the century was the sixth highest in the world [50], and the vast majority (58%) of clearing over the last three decades has occurred in Queensland. Successive policies have been introduced in Queensland since 1995 in an attempt to control deforestation [1,51,54,55], with amendments in 2007 said to have signaled the "end of broad-scale land clearing" in Australia [56,57]. The national downturn in forest loss since the 1990's has been attributed to state-level regulations on native forest clearing [58–60], but often with limited empirical basis [51,53,61].

Recent work has sought to empirically test the effect of forest policy interventions in Australia [1,52,55]. Evans [1] explored the relationship between deforestation rates and known macroeconomic, climatic and spatial drivers at a national scale. Marcos-Martinez and colleagues [52] conducted finer-scale analysis of these drivers, and highlighted the likely contribution of regulatory policy around 2008 in Queensland to forest transition in Australia. Rhodes and colleagues [62] identified an overall positive influence of regulation on forest cover in Queensland, using path analysis within a Bayesian framework. However, they also found evidence that highly threatened forest types were far less effectively protected than non-threatened forests. Simmons and colleagues used a synthetic control method to estimate the causal impact of regulatory policies on the rate of forest loss in Queensland [53]. Using similar datasets, the same authors applied matching techniques and a set of alternative counterfactuals to infer the causal impact of regulation on remnant (primary) forest loss in the Brigalow Belt region, a deforestation hotspot in Queensland [61]. Both studies identified a positive influence of regulation on forest cover, though effect size varied by region of interest, forest cover type (primary versus all forest ages) and counterfactual assumptions.

In this study, we adapt the bent-cable regression model [63–65] to evaluate the impact of regulatory policies on deforestation in Queensland relative to macroeconomic, institutional, biophysical and climatic drivers. The bent-cable model generalizes the broken-stick (piecewise-linear) model by providing a more realistic and flexible representation of system changes [63,64,66], yet has so far been under-utilised for policy evaluation [65,67]. In particular, suppose the policy introduced at time $t$ unequivocally results in a reduction in deforestation. Then, even after adjusting for driver effects, the temporal trend in deforestation must exhibit a higher rate of decline sometime after $t$ than the rate before time $t$. In reality, any observable impact of an environmental policy is rarely unequivocal. In such cases, the bend of an empirically fitted bent cable would provide quantitative clues to the presence and nature of a change in deforestation rate before and after the introduction of regulatory policies.

In the remainder of this paper, we first document recent deforestation rates in Queensland, and describe how historical and present-date drivers have contributed to forest loss over time. Second, we briefly describe the key policy changes that have occurred in our study region in the past three decades, and highlight which interventions we expect to have influenced the rate of deforestation beyond the effects of macroeconomic, climatic, biophysical and institutional variables. Third, we provide details on the data used to model deforestation over time, and justify our selection of covariates used in the statistical analysis. We then outline our bent-cable regression model specification, and report on the results of our analysis. Finally, we discuss the implications of our findings for the design and implementation of policies aiming to control deforestation in Australia and other forested nations which rely on agriculture and other commodity exports for economic development.

## Background

### Study region and policy context

Queensland is Australia's second-largest state, covering 170 million hectares (23% of the continent). Over 5 million hectares of forest has been cleared between 1988 and 2014, of which 31% was primary (remnant) forest (Figure 1) Agriculture is the primary land use, with extensive grazing making up 85% of the state by area, and cropping and other agricultural industries (excluding forestry) comprising just 2%. Beef is the state's most important agricultural commodity, contributing AUD 3.3 billion in 2013–14 [68] which is close to half of Australia's total production. The majority of land is leasehold (63%), with only 25% privately owned (freehold). Leasehold land in Queensland may be held in perpetuity or for a fixed term (1-100 years), with tenure issued for a specific purpose (e.g. agriculture). Deforestation occurs disproportionally on freehold land in Queensland [69] and in Australia as a whole [1].

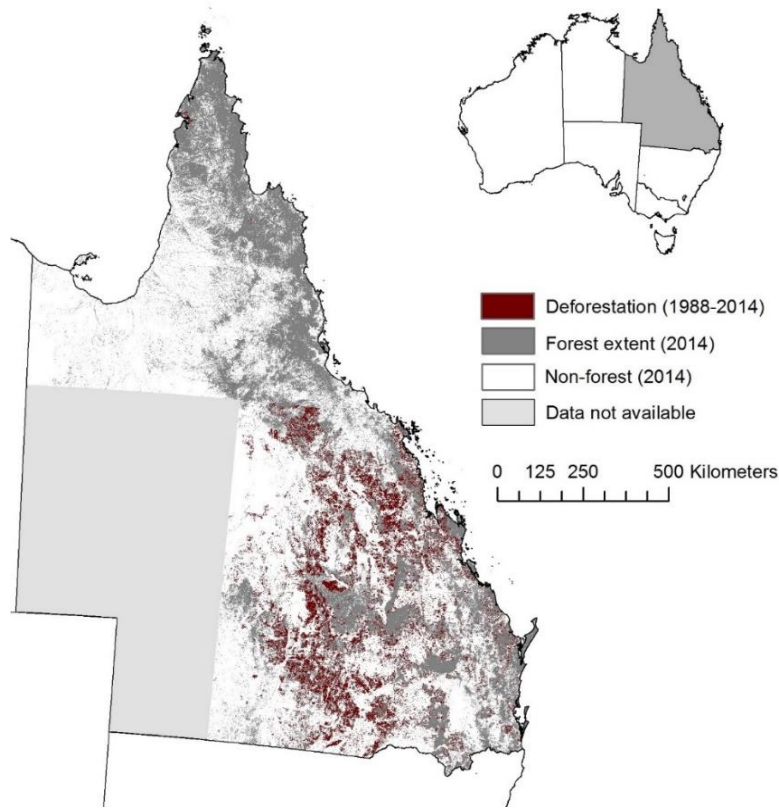

Figure 1. Deforestation events (including primary and regrowth deforestation) attributed to human intervention from 1988 − 2014. Data is sourced from the National Carbon Accounting System (NCAS) [70,71]

Queensland holds the largest remaining area of forest Australia, and the availability of cheap land suitable for agriculture has been key historical driver of deforestation [1,72]. Native vegetation such as Brigalow (*Acacia harpophylla*) and Mulga (*Acacia aneura*) grow vigorously, and are generally re-cleared in 15 year cycles to maintain suitable pasture [73,74]. Government land development schemes, access to cheap finance and tax concessions facilitated early agricultural expansion, but most of these incentives were removed by the 1980's [72,75]. Fluctuations in commodity prices, terms of trade, rainfall and regulatory controls on clearing are considered the key contemporary drivers of deforestation in Australia [1,52,58,72].

Deforestation in Queensland has occurred at an average rate of 181,000 hectares per year since 1988 (Figure 2). State-wide regulation of native vegetation clearing on leasehold land was first introduced under the *Land Act 1994* (Figure 2(a) [1,54]), whereas similar controls on freehold properties were enacted five years later under the *Vegetation Management Act 1999* (VMA, Figure 2(b)). The upturn in deforestation in 2000 has been attributed to "panic clearing" [55,57], where landholders cleared substantial amounts of vegetation prior to the VMA coming into effect in an effort to avoid regulation.

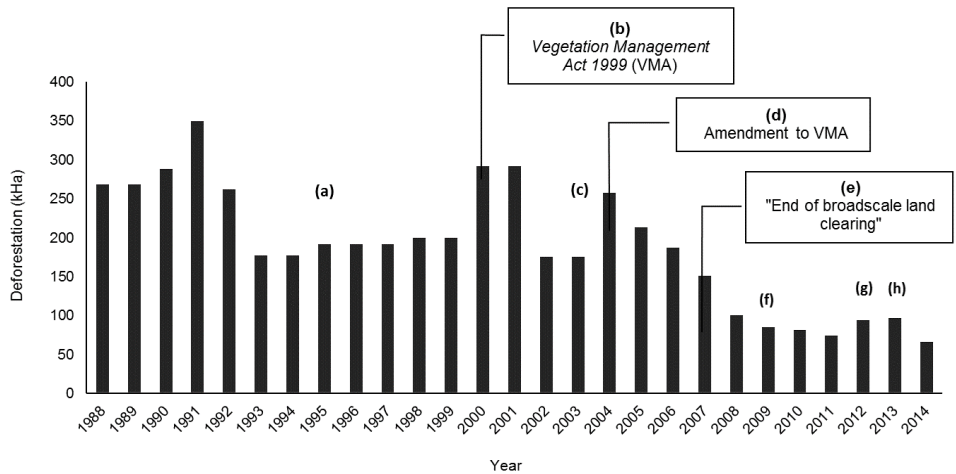

Figure 2. Annual rate of deforestation in Queensland from 1988-2014, and key legislative changes to deforestation reglation. The central piece of legislation is the *Vegetation Management Act 1999* (b), which came into force in 2000 to regulate native vegetation clearing on freehold land. Amendments to the VMA in 2004 (d) were said to led to the end of broad-scale land clearing by on 31[st] December 2006 (e). Other legislative changes corresponding to (a), (c) and (f) to (h) are described in Appendix 1.1. Note that policy changes occurring after 2014 are not considered in this current study due to data availability [70,71].

In 2004, a ballot was held for clearing permits totaling 500,000 ha which were all used or expired by 31 December 2006. This policy change is considered to have effectively reduced deforestation in Queensland to its lowest level in the last three decades. The relaxation of the VMA's regulatory controls by the conservative Queensland Government in 2013, and the subsequent increase in clearing rates [62,76] has added further weight to this claim.

Further amendments to the VMA were passed in 2018, following an election commitment by the Palaszczuk Queensland Government to restrengthen regulatory controls on deforestation [77]. However, our analysis only considers events up to 2014 due to data availability [70,71].

## Materials and Methods

### Deforestation data

We used national-scale spatial datasets on forest extent and human induced forest change [70,71] developed by the Australian Government as part of the National Carbon Accounting System [1,78,79]. The NCAS uses over 7,000 Landsat MSS, TM and ETM+ images to map forest extent and change from 1972 to 2014 at a 25 metre resolution across the Australian continent. Note that the NCAS classifies imagery according to a definition of 'forest' as vegetation with a minimum of 20% canopy cover, at least 2 metres high, and with a minimum area of 0.2 hectares.

Annual forest extent and deforestation data are not available within the NCAS until 2005. Prior to then, data are instead captured within multi-year epochs (instances in time), with some epochs (e.g 1972) containing data for five consecutive years [1,70,71]. We converted deforestation events contained within multi-year epochs into annual values by dividing the deforestation that occurred within that epoch by the number of years within the epoch [71] (Reddy, S, pers. comm.). For example, the 1972 epoch contains deforestation events for 1972 to 1977, hence the annual amount of deforestation in those years is the deforestation occurring in the 1972 epoch divided by five. To derive annual forest extent values, we simply assumed that the forest extent value in a multi-year epoch was equivalent to the corresponding annual values. For example, the area of forest extent in the 1988 epoch was assigned as the annual forest extent value for 1988 and 1989.

Preliminary modelling results suggested that an unreasonable amount of uncertainty in the model estimates was a consequence of (a) the lack of real observed data (as opposed to artificially imputed data) prior to 1988 and (b) the ambiguity in the definition of primary (remnant) forest extent. Therefore, for our final models, we discarded sparse data early in the time series by considering all deforestation events and forest extent [70,71] from 1988 to 2014 only. By "all deforestation events and forest extent" we did not differentiate between primary and regrowth forest as per Evans [1].

Using a national land use dataset [80], we excluded protected areas and areas of commercial forestry from the analysis, and so considered deforestation only where the land use was for residential and urban development, agriculture, grazing and mining. Local Government Areas (LGAs) were selected as spatial units of analysis due to their relatively small size and alignment with catchment boundaries, which have previously been used spatially differentiate native vegetation regulations in Queensland [1]. State-level native vegetation legislation does not prevent a local planning restrictions on deforestation in an LGA, however the State law prevails in the event of inconsistencies [81]. We used the 'raster' package [82] in R Statistical software [83] to summarise forest extent and deforestation data at the LGA level.

**Macroeconomic data**

We extracted key national level indicators including the annual percentage growth in gross domestic product (GDP, USD in real terms), the value added by agriculture to GDP, and terms of trade [84]. Oil supply constraints are associated with changes in the rate of deforestation [85], hence we considered the annual imported crude oil price (USD/barrel) [86]. We used national scale agricultural commodity statistics [87] as local surveys are not conducted annually.

**Climate data**

Rainfall is a key driver of rural land use decisions [51,88], and is expected to affect deforestation rates independently of broader macroeconomic variables [58]. Queensland regularly experiences periods of drought, and these regional climatic extremes can influence observed deforestation rates due to changes in the availability and quality of fodder [72]. Spatiotemporal variation in temperature and rainfall is also linked to the relative suitability of land for grazing and cropping [52]. We obtained spatial data for key climatic variables from the Australian Bureau of Meteorology [89]. Rainfall, vapour pressure ('humidity') and temperature were sourced as monthly averages [90–92]. Quantities were calculated within each LGA polygon for each year from 1990 to 2014 using the Python Rasterstats package [93].

**Biophysical data**

We derived slope and elevation data from a 1 second Digital Surface Model [94] to account for the influence of topography on agricultural productivity and deforestation [52,95,96]. Spatial variation in vegetation productivity was accounted for using the normalized difference vegetation index [97]. The mean, median and standard deviation was calculated for each LGA using the Rasterstats package [93].

**Land tenure**

We used the most recent national land tenure dataset [98] to determine the proportion of each LGA held in leasehold, freehold or public tenure. We derived a spatial covariate $L_i$ (Figure 3) to represent the extent to which an LGA $i$ is held in leasehold or freehold tenure, according to:

$$L_i = \log \frac{(\% \, freehold + 0.01)}{(\% \, leasehold + 0.01)}$$

Table 1 provides a comprehensive summary of all variables considered in our analysis during covariate selection, and the covariates selected for inclusion in our final model. Details of the covariate selection process is provided in the Appendix 1.2.

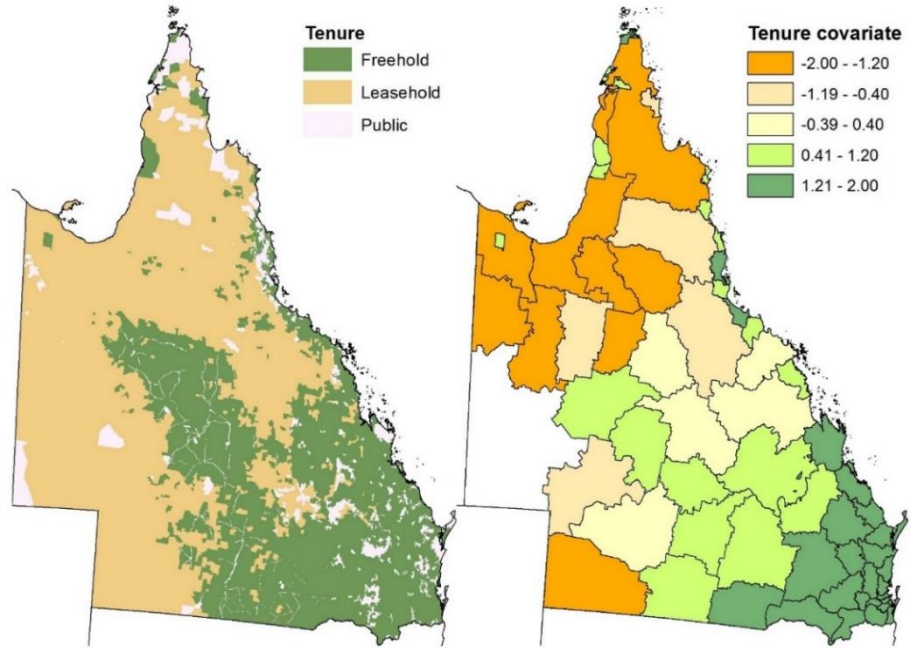

Figure 3. Land tenure in Queensland (left, [97]) and corresponding tenure covariate $L_i$ calculated for each Local Government Areas (LGAs, n = 74 (total), n = 50 (included in regression analysis))

Table 1. Key variables considered in our regression analysis

| Name | Type | Description | Spatial resolution and units | | Temporal resolution | | Contained in final model? | Data source |
|---|---|---|---|---|---|---|---|---|
| | | | Original | Final | Original | Final | | |
| *Response variable* | | | | | | | | |
| Area of deforestation | Environmental | Extent of total forest loss attributed to human intervention | 0.0002 degrees; 25m | Hectares, calculated for each LGA | Available for epochs: 1972, 1977, 1980, 1985, 1989, 1991, 1992, 1995, 1998, 2000, 2002, 2004, annually from 2005 to 2014 | 1988 to 2014 (annual) | - | Australian Department of the Environment (2015) |
| Area of forest extent | | Extent of vegetation classified as 'forest' (excluding protected areas and commercial forestry) | | | | | - | Australian Department of the Environment (2016) |
| Proportion of deforestation ($y_{it}$) | | Area of deforestation relative to the area of forest available to clear in LGA I and year t | LGA; % | | | | Yes | Derived |
| *Explanatory variables* | | | | | | | | |
| Rainfall | Climatic | Total annual rainfall | | mm, calculated for each LGA | | 1988 to 2014 (annual) | - | Bureau of Meteorology (2017a) |
| | | Minimum of mean monthly minimum observations | | | | | | |
| | | Maximum of mean monthly minimum observations | | | | | | |
| Vapour pressure ('humidity') | | Mean of mean monthly 9am estimates | 0.05 degrees; 5km | hPa, calculated for each LGA | 1900 to current | 1988 to 2014 (annual) | Yes | Bureau of Meteorology (2017c) |
| | | Mean of mean monthly minimum observations | | | | | - | |
| Temperature | | Minimum of mean monthly minimum observations | | degrees celsius, calculated for each LGA | | 1988 to 2014 (annual) | - | Bureau of Meteorology (2017b) |
| | | Mean of mean monthly maximum observations | | | | | - | |
| | | Maximum of mean monthly maximum observations | | | | | - | |
| Land tenure | Institutional | Extent of land tenures across Australia | Polygon | 25m raster | - | | - | Geoscience Australia (1993) |
| Tenure covariate ($L_i$) | | Log ratio of % freehold to % leasehold in each LGA | Index from -2 to 2. Calculated for each LGA | | | | Spatial adjacency only | Derived |

| Name | Type | Description | Spatial resolution and units | | Temporal resolution | | Contained in final model? | Data source |
|---|---|---|---|---|---|---|---|---|
| | | | Original | Final | Original | Final | | |
| *Explanatory variables* | | | | | | | | |
| Slope | | Mean, median, minimum, maximum | 1 second (approx 30m) | m, calculated for each LGA | - | | - | Geoscience Australia (2017) |
| Elevation | | | | | | | Yes (mean) | |
| NDVI | Biophysical | Mean of long term mean (estimate of productivity) | 10km | Calculated for each LGA | 1981 to 2011 | - | - | Tucker et al. (2014) |
| | | Standard deviation of long term mean (estimate of land cover variability ) | | | | | | |
| | | Mean of long term standard deviation (estimate of production variability over time) | | | | | | |
| % GDP growth (annual) | | % growth in the value of all goods and services produced in a given year | National, % | | | | Yes | |
| % value added to GDP by agriculture | | Net output of agriculture after adding up all outputs and subtracting intermediate inputs | National, % of GDP | | 1972 to 2014 | | - | The World Bank (2015) |
| Inflation | | Rate of price change in the economy as a whole | National, % | | | | | |
| Oil price | Macroeconomic | Annual average imported crude oil price | International, USD/barrel | | 1968 to 2015 | 1988 to 2014 (annual) | Yes | U.S. Energy Information Administration (2015) |
| Terms of trade (ToT) | | Ratio of total export prices to import prices | National | | | | | |
| Farmer's terms of trade (FToT) | | Ratio of prices received by farmers to prices paid by farmers | National | | 1972 to 2015 | | - | ABARES (2015) |
| Gross value of agricultural exports | | Gross value of farm production, value of cereal exports, meat exports, wool exports, total exports | National, AUD | | | | | |

**Model specification**

We developed a holistic model of deforestation trends in Queensland to model LGAs (n = 74) as the spatial units of analysis. Forest data were not available in 6 LGAs in the western part of the state, and we discarded a further 18 LGAs due to insufficient data (Appendix 1.3, Table A1.3), leaving 50 LGAs remaining to be analysed. Our response variable $y_{it}$ is calculated as the relative proportion of deforestation in LGA $i$ at time step $t$, then log-transformed twice to reduce skewness:

$$y_{it} = \log\left\{-\log \frac{(\text{Area of deforestation})_{it}}{(\text{Area of forest extent})_{it}}\right\}$$

The total forest extent in LGA $i$ may decrease, increase or remain unchanged between sequential years, since the NCAS captures forest as it is cleared, regrown and re-cleared. This means that $y_{it}$ similarly may decrease, increase or remain unchanged between time steps. Our specification of $y_{it}$ controls for the area of forest available to clear [1,72], such that if no forest is present in time $t$, no deforestation may occur.

Our hierarchical (multilevel) regression framework collectively models deforestation across Queensland (i.e the population level) by drawing on LGA-specific bent-cable model estimates. The bent-cable function comprises two linear segments (the incoming and outgoing phase), connected by a quadratic bend (Figure 4). The linear segments are parametrized by an intercept $\alpha_{0_i}$, incoming slope $\alpha_{1_i}$, and outgoing slope $\alpha_{1_i} + \alpha_{2_i}$.

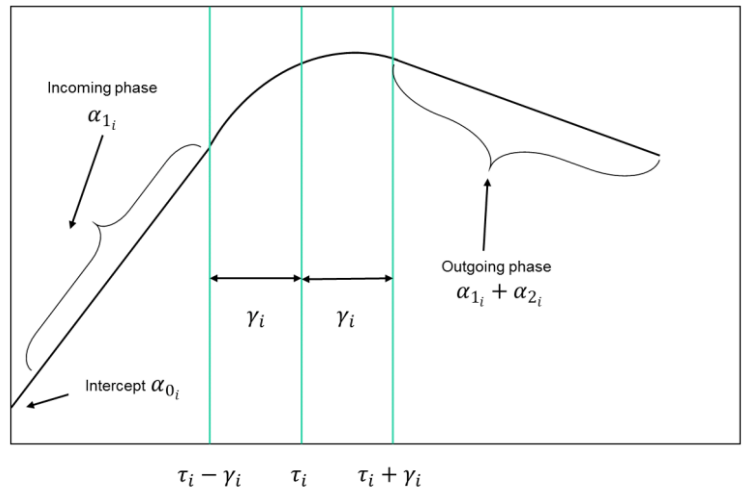

Figure 4. General depiction of the bent-cable function, adapted from [65]

Our modeling framework below accounts for spatial correlation among LGAs, and the longitudinal nature of the data (i.e. each LGA has its own time series data):

$$i=1,\ldots,n_{\text{LGA}}(=50)$$
$$t=1,\ldots,n_T(=27)$$
$b_0, \boldsymbol{b}_1, \boldsymbol{b}_2, b_{15}, b_{23}$: population-level driver slope parameters (fixed effects)
$a_1, a_2, \mathcal{T}, \ell_\gamma$: population-level bent-cable parameters (fixed effects)
$v, \sigma_\tau, \sigma_\gamma, \sigma_1, \sigma_2, \sigma_{10}, \sigma_{20}$: population-level dispersion parameters (fixed effects)
$$\boldsymbol{E}_t = (\text{macroeconomic data vector})_t$$
$$\boldsymbol{C}_{it} = (\text{climate data vector})_{it}$$

Level 1:

$$\begin{cases} y_{it} = (b_0 + \beta_{10i}) + b_{15}L_i + b_{23}(\text{elevation})_i (\text{i.e., spatial-only covariates}) + \\ \qquad \beta_{20t} + \boldsymbol{b}_1'\boldsymbol{E}_t (\text{i.e., temporal-only covariates}) + \\ \qquad \mu_{it}(\text{i.e., spatio-temporal term}) + \epsilon_{it} \\ \qquad \epsilon_{it} \sim N(0, v^2) \\ \mu_{it} = \boldsymbol{b}_2'\boldsymbol{C}_{it} + \alpha_{1i}t + \alpha_{2i}\dfrac{(t-\tau_i+\gamma_i)^2}{4\gamma_i}1\{|t-\tau_i| \le \gamma_i\} + \\ \qquad \alpha_{2i}(t-\tau_i)1\{t > \tau_i + \gamma_i\} \\ (\text{i.e., spatiotemporal covariates and LGA-specific bent cable}) \end{cases}$$

Level 2:

$$\begin{cases} \alpha_{1i} \sim N(a_1, \sigma_1^2) \\ \alpha_{2i} \sim N(a_2, \sigma_2^2) \\ \tau_i \sim N(\mathcal{T}, \sigma_\tau^2) \\ \log\gamma_i \sim N(l_\gamma, \sigma_\gamma^2) \\ \beta_{10i} \sim \text{CAR}(\sigma_{10}^2) \text{ (i.e., random LGA-specific deviation} \\ \qquad \text{from population intercept } b_0) \\ \beta_{20t} \sim N(0, \sigma_{20}^2) \text{ (i.e., random year-specific deviation} \\ \qquad \text{from population intercept } b_0) \end{cases}$$

where "CAR" stands for the "conditional autoregressive" spatial correlation structure that we assume among LGAs. CAR is the spatial analogy of the lag-one autoregressive (AR(1)) temporal structure. It stipulates that given any LGAs $i$ and $j$, their response values $y_{it}$ and $y_{jt}$ mutually influence each other only if $i$ and $j$ share borders [64,67,99]. Note that Khan and colleagues [67] also employ the CAR structure under a spatial-longitudinal bent-cable framework to evaluate the impact of the Montréal Protocol on the reduction of atmospheric chlorofluorocarbons. However, they only consider 8 spatial units under an ambiguous definition of spatial adjacency.

Statistical inference is made under the Bayesian paradigm. The posterior distribution from which Bayesian estimates are derived requires prior distributions to be specified for all population-level parameters (see Appendix 1.4 for details). Models are implemented in R using the R2WinBUGS package [100]. WinBUGS numerically approximates the posterior distribution [101–103].

To explore relevance of land tenure on spatial adjacency, we draw on the approach taken by Earnest and colleagues [99] and define spatial weighting according to:

$$w_{ij} = \frac{\mathbf{1}\{i, j \text{ share border}\}}{\mid L_i - L_j \mid + 0.00001}$$

Such weighting stipulates that neighbouring LGAs $i$ and $j$ would influence each other more than neighbouring LGAs $i'$ and $j'$ if $L_i$ and $L_j$ were more similar than $L_{i'}$ and $L_{j'}$.

We consider two key variations to our model framework described above, to explore:

a) Whether the "bend" occurs around the year 2000 at the introduction of the original VMA, or around 2007 after the VMA amendments came into effect; and
b) If weighting spatial adjacency by land tenure similarity makes a difference to model fit, as compared to unweighted spatial adjacency

Other minor model variations were considered to address model goodness-of-fit (Appendix 1.4)

## Results

Having tested the influence of over 20 covariates on deforestation behavior, our model selection procedure revealed that annual % growth in GDP and year were the only clear predictor variables of LGA-specific deforestation in a combined regression model. In other words, GDP can be regarded as a clear driver of deforestation at the LGA level even in the presence of other predictor variables, while other macroeconomic and climate predictors are confounded with year. Land tenure and elevation were also not statistically important drivers. Appendix 1.5 provides justification for discarding these model covariates.

Our model shows strong evidence of spatial contagion in deforestation (Figure 5), and that this effect is strengthened by spatial weighting which accounts for land tenure similarity between each LGA (median deviance ~ -1500 with weighting, ~ -930 without weighting; a smaller deviance indicates better goodness-of-fit; Appendix 1.6, see also Figure A1.2)

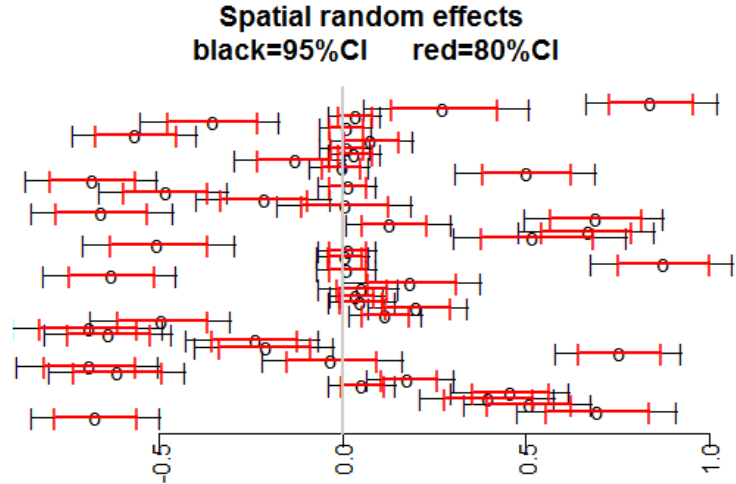

Figure 5. Estimates of $\beta_{10i}$ ('o') and corresponding Bayesian confidence intervals (CIs) at the 80% level (black) and 95% level (red). Strong evidence of spatial contagion is suggested by the large number of CIs which exclude 0.

Our results also suggest that deforestation in Queensland exhibits a population-level (state-wide) "bend" mostly between 2000 and 2007, consistent with expectations [52,61] (Figure 6). However, extreme variation is inherent in the deforestation time series (detrended by removing LGA-level driver effects from $y_{it}$) between and within LGAs, so that the signal for a state-wide bend is not particularly strong.

We found that LGA-level bent-cable estimates (Figure 7) can differ substantially. For some LGAs, deforestation increased over time to around 2000 then decreased (e.g Goondiwindi, Figure 7), as predicted by the timing of policy changes (Figure 2). Other LGAs followed the completely opposite trend (e.g Redland) and displayed an upturn in deforestation around 2007 which lasted until the end of the time series. Deforestation in some LGAs remained consistently high relative to the population-level trend (e.g Hope Vale) from 1990 to 2014, or followed a general declining trend (e.g Central Highlands). In yet more LGAs, deforestation peaked around 2007 and subsequently declined (e.g. Quilpie).

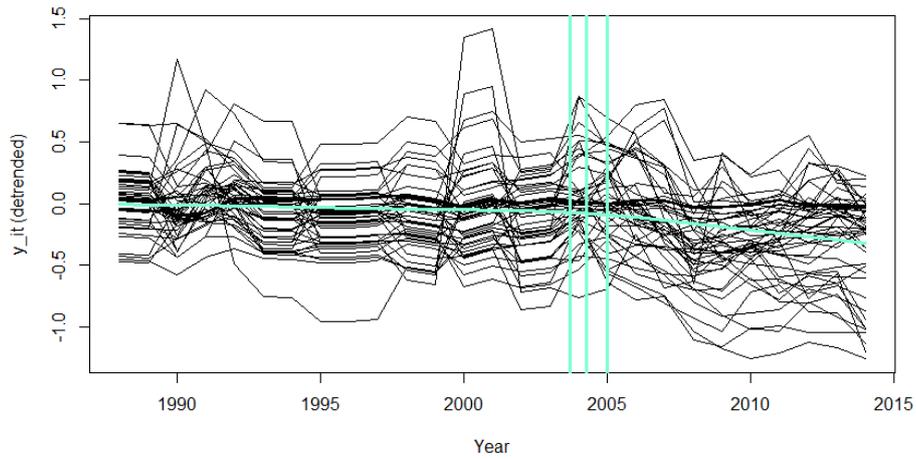

Figure 6. LGA-level time series $y_{it}$ ($n = 50$) after detrending (black) and fitted population-level bent cable (green). "Detrending" refers to removing LGA-level driver effects from the $y_{it}$ data. Vertical lines delimit the estimated bend (start, middle, and end of transition phase)

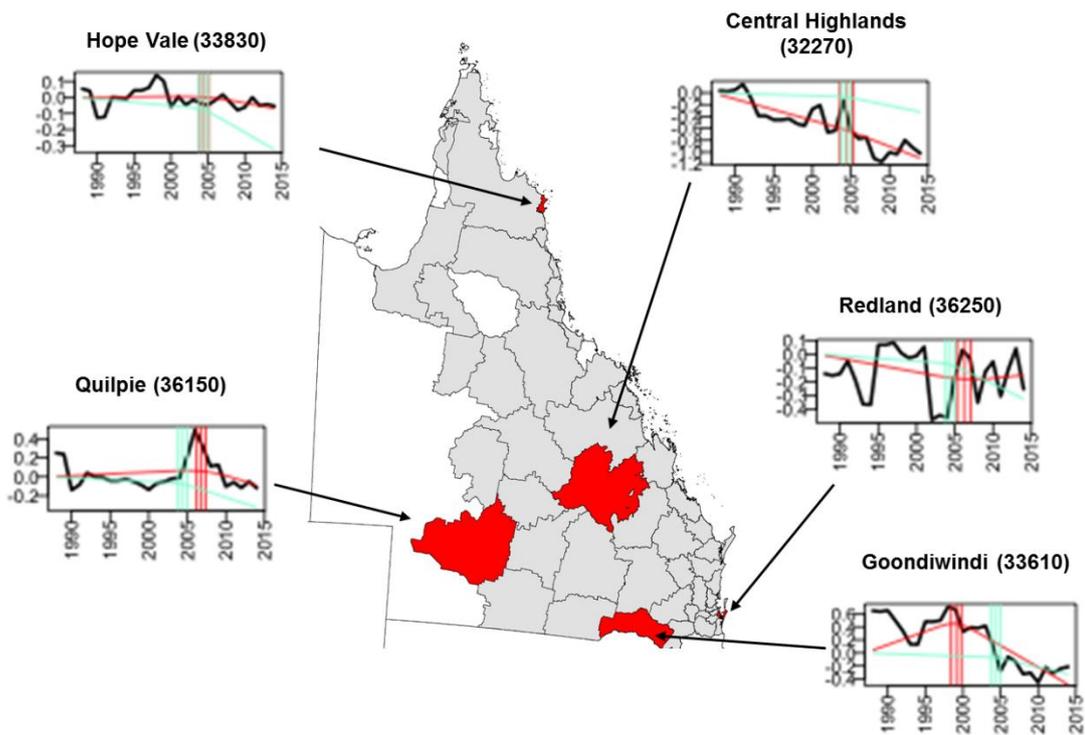

Figure 7. Examples of LGA-level time series $y_{it}$ ($n = 50$) after detrending (black), with LGA-level bent-cable estimate (red) and fitted population-level bent cable (green). See also Appendix 1.8, Figure A1.3.

## Discussion

We found that the spatially explicit bent-cable model is a promising technique for detecting a policy effect simultaneously at the local government area (LGA) level and state level while controlling for driver variables; and there is some evidence of policy-induced shifts in the deforestation rate as expected in 2000 with the introduction of the original VMA, and with the subsequent amendments in 2007. However, due to an inadequate number of data points (LGAs) under extreme spatial variation in trends in, and responses to, deforestation within and between local government areas, the overall level of statistical confidence may not be overwhelmingly high. Nevertheless, our analysis quantitatively supports that while deforestation in some LGAs increased in 2000 and decreased in 2007 as predicted by the policy changes, other LGAs followed the completely opposite trend, while deforestation remained either consistently high or low throughout the time series (1990 to 2014). Although we used different modelling techniques, our findings broadly agree with those of Simmons and colleagues [53,55], who similarly identified strong spatial variation in the contribution of policy timing and biophysical factors on deforestation rates [55] and a positive but weak effect of deforestation regulation at the state scale [53].

A core challenge with any statistical analysis is for there to be sufficient data points which may permit detection of the effect of policy introduction amidst substantial variation inherent in the data. This is understandably difficult to achieve at the national/state scale, especially with limited temporal continuity in data points. Expanding our analysis to the national scale could improve the capacity to detect any effect of policy interventions on deforestation, but at the expense of additional complexities given the considerable variation in timing and scope of regulatory policies across different Australian states [1]. It should be noted that little improvement in our capacity to estimate a "bend" (that could indicate a policy effect) would be expected by considering the data at a finer spatial resolution than LGA (for example in an equal-area grid), as the total amount of information per region in a grid would be offset by the substantially higher spatial correlation across gridded regions.

Using LGAs as a spatial unit of analysis provides the benefit of being able to investigate plausible explanations for regional deforestation trends using data captured within our tested variables (Table 1), or additional contextual information we have far been unable to quantify. For example, the Goondiwindi LGA is held entirely in freehold tenure (Table A1.2, excluding publicly managed protected areas). The LGA-specific "bend" around 2000 is consistent with the "panic clearing" phenomenon documented in 2000 [1,57], where deforestation spiked prior to the introduction of the original VMA which imposed restrictions on deforestation on freehold tenure for the first time. Hope Value is an Aboriginal Shire Council with a population of only 1,125 people [104], and several exemptions exist under the VMA to enable economic development on Indigenous land [105]. The upturn in deforestation around 2007 in the Redland LGA (Figure 7) can be explained by a spike in urban and residential development[5]. Exemptions under the VMA allow for native vegetation clearing to proceed for urban development (e.g. residential, industrial, sporting, recreational or commercial) in a regional ecosystem [108] that is listed as 'least concern' [105][6]. Land tenure in the Quilpie LGA is equally split between freehold and leasehold (Table A2.2), and is dominated by Mulga (*Acacia aneura*) which is regularly "pushed" to maintain pasture, and to provide emergency feed for cattle during periods of drought [73,74,109].

Our present analysis sought to quantitatively establish evidence for the impact of regulatory policies aimed at reducing deforestation in Queensland, which are widely regarded to have led to a state-wide and national decline in deforestation around 2007 [1,57,58,76]. Previous work has emphasized the overarching influence of macroeconomic variables [26–28], agricultural commodity prices and terms of trade [38–40,58], rainfall and temperature [52,58], biophysical variables [52,110] and institutions [9,41–44] in driving deforestation. We tested representative variables from each of these groups (Table 1), and our modelling framework (which combined driver variables with year as covariates within a single regression) revealed that most of these variables were highly correlated or provided no extra information in addition to year (Appendix 1.2), or did not improve model goodness-of-fit (Appendix 1.5).

The Australian Government uses a linear regression model (with farmers' terms of trade as a single covariate) to predict future deforestation up to 2035. This model informs national land use, land use change and forestry (LULUCF) emissions projections for reporting under the UN Framework Convention on Climate Change (UNFCCC) [58,111]. Although this simple model may be adequate for making a national-level prediction of future deforestation trends, our spatially explicit analysis suggests that the assertion of a "drop in land clearing activity from 2007 onwards" [111] based on a single-variable regression is premature. In future work, our quantitative evaluation approach could be strengthened by including spatially explicit information not often considered in deforestation modelling, such as broad vegetation groups [62,108], primary land use [80] and profitability [112], and social data which may reveal landholder compliance behaviours [55,113].

---

[5] The koala is a charismatic species[2] which was once abundant throughout the Redland region, but has suffered an estimated 80.3% decline in population densities between 1996 and 2014, primarily due to urban development [106,107].

[6] Note that the Queensland, New South Wales and Australian Capital Territory populations of the koala (*Phascolarctos cinereus*) were listed as Vulnerable under the *Environment Protection and Biodiversity Conservation Act 1999* (EPBC Act) in 2012, and so would not have substantially influenced the observed deforestation trend in Redland.

Our analysis used the 2015 version of human induced forest change data as developed by the Australian Government for the NCAS [70,71]. The Queensland Government also runs a system for detecting and analyzing land use change under the Statewide Landcover and Trees Study (SLATS) [69]. Recent work has identified substantial differences in the amount of deforestation estimated by the NCAS and SLATS systems [114], largely due to an inconsistent definition of 'forest'[7] [115]. To determine whether regulatory policies introduced in Queensland have affected local deforestation, the SLATS data would be more fit for purpose. However, collectively modelling deforestation at the national scale would still require use of the NCAS system given the substantial differences in deforestation accounting methods used in each Australian state [114].

## Conclusions

Deforestation in Australia occurs at a rate and scale which is on par with well-known global deforestation hotspots [30,31], despite the presence of strong governance [48,49] and the introduction of a series of policies aimed to control the clearing of native vegetation over the past four decades [1]. Given the substantial and widespread impacts of deforestation on biodiversity, ecosystem service provision, and climate regulation, a greater understanding of the efficacy of policy interventions in regions where deforestation may be assumed to be effectively "controlled" therefore warrants further attention.

Although our population-level findings should be considered preliminary given the substantial "unexplained" variation in LGA-specific deforestation trends, our results nonetheless suggest a degree of caution should be taken before proclaiming the effectiveness of regulatory policies prior to conducting an *ex-post* evaluation. Data limitations and methodological challenges can impede the evaluation of policy impact in the presence of confounding variables, time-lags and misalignment of spatial and temporal data. Such challenges in evaluation must be overcome to ensure that policies effectively deliver environmental outcomes as anticipated.

---

[7] The NCAS adopts a Kyoto definition of 'forest' where canopy cover must be at least 20%. However, much of the historical and recent native vegetation clearing in Queensland is 'sparse woody vegetation' (e.g Mulga and Brigalow ecosystems) which does not always meet the 20% canopy cover threshold. The NCAS is designed to monitor forest loss for the purpose of reporting Australia's emissions under the UNFCCC, whereas the SLATS program is used specifically to monitor for compliance under the Queensland VMA.

## Acknowledgments

This research was conducted with the support of funding from the Australian Government's National Environmental Research Program. MCE was supported by an Australian Postgraduate Award and a CSIRO Climate Adaptation Flagship Top-up Scholarship. We are grateful to Peter Scarth for his assistance in extracting climate and topographic data, and to Shanti Reddy for providing advice on the forest change and extent data.

## References

1. Evans MC. Deforestation in Australia: drivers, trends and policy responses. Pacific Conservation Biology. 2016;22: 130–150.

2. Kirby KR, Laurance WF, Albernaz AK, Schroth G, Fearnside PM, Bergen S, et al. The future of deforestation in the Brazilian Amazon. Futures. 2006;38: 432–453.

3. Lambin EF, Turner BL, Geist HJ, Agbola SB, Angelsen A, Bruce JW, et al. The causes of land-use and land-cover change: moving beyond the myths. Global Environmental Change. 2001;11: 261–269.

4. Vitousek PM, Mooney HA, Lubchenco J, Melillo JM. Human Domination of Earth's Ecosystems. Science. 1997;277: 494–499. doi:10.1126/science.277.5325.494

5. Andam KS, Ferraro PJ, Pfaff A, Sanchez-Azofeifa GA, Robalino JA. Measuring the effectiveness of protected area networks in reducing deforestation. Proceedings of the National Academy of Sciences of the United States of America. 2008;105: 16089–94.

6. Gaveau DLA, Epting J, Lyne O, Linkie M, Kumara I, Kanninen M, et al. Evaluating whether protected areas reduce tropical deforestation in Sumatra. Journal of Biogeography. 2009;36: 2165–2175.

7. Joppa LN, Pfaff A. Global protected area impacts. Proceedings of the Royal Society B: Biological Sciences. 2011;278: 1633–1638. doi:10.1098/rspb.2010.1713

8. Ferraro PJ, Hanauer MM, Miteva DA, Canavire-Bacarreza GJ, Pattanayak SK, Sims KRE. More strictly protected areas are not necessarily more protective: evidence from Bolivia, Costa Rica,

Indonesia, and Thailand. Environmental Research Letters. 2013;8: 025011. doi:10.1088/1748-9326/8/2/025011

9. Nolte C, Agrawal A, Silvius KM, Soares-Filho BS. Governance regime and location influence avoided deforestation success of protected areas in the Brazilian Amazon. Proceedings of the National Academy of Sciences. 2013;110: 4956–4961. doi:10.1073/pnas.1214786110

10. Gaveau DL a., Curran LM, Paoli GD, Carlson KM, Wells P, Besse-Rimba a., et al. Examining protected area effectiveness in Sumatra: importance of regulations governing unprotected lands. Conservation Letters. 2012; no-no. doi:10.1111/j.1755-263X.2012.00220.x

11. Hargrave J, Kis-Katos K. Economic Causes of Deforestation in the Brazilian Amazon: A Panel Data Analysis for the 2000s. Environ Resource Econ. 2013;54: 471–494. doi:10.1007/s10640-012-9610-2

12. Arima EY, Barreto P, Araújo E, Soares-Filho B. Public policies can reduce tropical deforestation: Lessons and challenges from Brazil. Land Use Policy. 2014;41: 465–473. doi:10.1016/j.landusepol.2014.06.026

13. Nepstad D, McGrath D, Stickler C, Alencar A, Azevedo A, Swette B, et al. Slowing Amazon deforestation through public policy and interventions in beef and soy supply chains. Science. 2014;344: 1118–1123. doi:10.1126/science.1248525

14. Soares-Filho B, Rajão R, Macedo M, Carneiro A, Costa W, Coe M, et al. Cracking Brazil's Forest Code. Science. 2014;344: 363–364. doi:10.1126/science.1246663

15. Busch J, Ferretti-Gallon K, Engelmann J, Wright M, Austin KG, Stolle F, et al. Reductions in emissions from deforestation from Indonesia's moratorium on new oil palm, timber, and logging concessions. Proceedings of the National Academy of Sciences. 2015;112: 1328–1333. doi:10.1073/pnas.1412514112

16. SáNchez-Azofeifa GA, Pfaff A, Robalino JA, Boomhower JP. Costa Rica's Payment for Environmental Services Program: Intention, Implementation, and Impact. Conservation Biology. 2007;21: 1165–1173. doi:10.1111/j.1523-1739.2007.00751.x

17. Liu J, Li S, Ouyang Z, Tam C, Chen X. Ecological and socioeconomic effects of China's policies for ecosystem services. PNAS. 2008;105: 9477–9482. doi:10.1073/pnas.0706436105

18. Arriagada RA, Ferraro PJ, Sills EO, Pattanayak SK, Cordero-Sancho S. Do Payments for Environmental Services Affect Forest Cover? A Farm-Level Evaluation from Costa Rica. Land Economics. 2012;88: 382–399. doi:10.3368/le.88.2.382

19. Meyfroidt P, Lambin EF. Forest transition in Vietnam and displacement of deforestation abroad. PNAS. 2009;106: 16139–16144. doi:10.1073/pnas.0904942106

20. Brandt JS, Nolte C, Agrawal A. Deforestation and timber production in Congo after implementation of sustainable forest management policy. Land Use Policy. 2016;52: 15–22. doi:10.1016/j.landusepol.2015.11.028

21. Ferraro PJ. Counterfactual thinking and impact evaluation in environmental policy. New Directions for Evaluation. 2009;2009: 75–84. doi:10.1002/ev.297

22. Miteva DA, Pattanayak SK, Ferraro PJ. Evaluation of biodiversity policy instruments: what works and what doesn't? Oxford Review of Economic Policy. 2012;28: 69–92. doi:10.1093/oxrep/grs009

23. Ferraro PJ, Hanauer MM. Advances in Measuring the Environmental and Social Impacts of Environmental Programs. Annual Review of Environment and Resources. 2014;39: 495–517. doi:10.1146/annurev-environ-101813-013230

24. Ferraro PJ, Hanauer MM, Sims KRE. Conditions associated with protected area success in conservation and poverty reduction. Proceedings of the National Academy of Sciences. 2011;108: 13913–13918. doi:10.1073/pnas.1011529108

25. Nolte C, le Polain de Waroux Y, Munger J, Reis TNP, Lambin EF. Conditions influencing the adoption of effective anti-deforestation policies in South America's commodity frontiers. Global Environmental Change. 2017;43: 1–14. doi:10.1016/j.gloenvcha.2017.01.001

26. Angelsen A, Kaimowitz D. Rethinking the Causes of Deforestation: Lessons from Economic Models. The World Bank Research Observer. 1999;14: 73–98. doi:10.1093/wbro/14.1.73

27. Geist HJ, Lambin EF. Proximate Causes and Underlying Driving Forces of Tropical Deforestation. BioScience. 2002;52: 143–143.

28. Angelsen A. Policies for reduced deforestation and their impact on agricultural production. Proceedings of the National Academy of Sciences of the United States of America. 2010;107: 19639–44.

29. Bartel RL. Satellite Imagery and Land Clearance Legislation: A Picture of Regulatory Efficacy? Australasian Journal of Natural Resources Law and Policy. 2004;9: 1–31.

30. Hansen MC, Potapov P V, Moore R, Hancher M, Turubanova S a, Tyukavina A, et al. High-resolution global maps of 21st-century forest cover change. Science (New York, NY). 2013;342: 850–3. doi:10.1126/science.1244693

31. Keenan RJ, Reams GA, Achard F, de Freitas J V., Grainger A, Lindquist E. Dynamics of global forest area: Results from the FAO Global Forest Resources Assessment 2015. Forest Ecology and Management. 2015;352: 9–20. doi:10.1016/j.foreco.2015.06.014

32. Laurance WF. Reflections on the tropical deforestation crisis. Biological Conservation. 1999;91: 109–117. doi:10.1016/S0006-3207(99)00088-9

33. Laurance WF, Useche DC, Rendeiro J, Kalka M, Bradshaw CJ a, Sloan SP, et al. Averting biodiversity collapse in tropical forest protected areas. Nature. 2012;489: 290–4. doi:10.1038/nature11318

34. Laurance WF, Sayer J, Cassman KG. Agricultural expansion and its impacts on tropical nature. Trends in Ecology & Evolution. 2014;29: 107–116. doi:10.1016/j.tree.2013.12.001

35. Jha S, Bawa KS. Population Growth, Human Development, and Deforestation in Biodiversity Hotspots. Conservation Biology. 2006;20: 906–912.

36. DeFries RS, Rudel T, Uriarte M, Hansen M. Deforestation driven by urban population growth and agricultural trade in the twenty-first century. Nature Geoscience. 2010;3: 178–181.

37. Laurance WF. The Future of the Brazilian Amazon. Science. 2001;291: 438–439. doi:10.1126/science.291.5503.438

38. Barbier EB, Rauscher M. Trade, tropical deforestation and policy interventions. Environ Resource Econ. 1994;4: 75–90. doi:10.1007/BF00691933

39. Ewers RM, Laurance WF, Souza CM. Temporal fluctuations in Amazonian deforestation rates. Environmental Conservation. 2008;35: 303–310.

40. Richards PD, Myers RJ, Swinton SM, Walker RT. Exchange rates, soybean supply response, and deforestation in South America. Global Environmental Change. 2012;22: 454–462. doi:10.1016/j.gloenvcha.2012.01.004

41. Fearnside PM. Land-Tenure Issues as Factors in Environmental Destruction in Brazilian Amazonia: The Case of Southern Pará. World Development. 2001;29: 1361–1372. doi:10.1016/S0305-750X(01)00039-0

42. Messina JP, Walsh SJ, Mena CF, Delamater PL. Land tenure and deforestation patterns in the Ecuadorian Amazon: Conflicts in land conservation in frontier settings. Applied Geography. 2006;26: 113–128. doi:10.1016/j.apgeog.2005.11.003

43. Dolisca F, McDaniel JM, Teeter LD, Jolly CM. Land tenure, population pressure, and deforestation in Haiti: The case of Forêt des Pins Reserve. Journal of Forest Economics. 2007;13: 277–289. doi:10.1016/j.jfe.2007.02.006

44. Robinson BE, Holland MB, Naughton-Treves L. Does secure land tenure save forests? A meta-analysis of the relationship between land tenure and tropical deforestation. Global Environmental Change. doi:10.1016/j.gloenvcha.2013.05.012

45. le Polain de Waroux Y, Garrett RD, Heilmayr R, Lambin EF. Land-use policies and corporate investments in agriculture in the Gran Chaco and

Chiquitano. Proc Natl Acad Sci U S A. 2016;113: 4021–4026. doi:10.1073/pnas.1602646113

46. Busch J, Ferretti-Gallon K. What Drives Deforestation and What Stops It? A Meta-Analysis. Rev Environ Econ Policy. 2017;11: 3–23. doi:10.1093/reep/rew013

47. Hansen MC, Stehman S V, Potapov P V. Quantification of global gross forest cover loss. Proceedings of the National Academy of Sciences. 2010; doi:10.1073/pnas.0912668107

48. Deacon RT. Deforestation and the Rule of Law in a Cross-Section of Countries. Land Economics. 1994;70: 414–430.

49. Bhattarai M, Hammig M. Governance, economic policy, and the environmental Kuznets curve for natural tropical forests. Environment and Development Economics. 2004;9: 367–382.

50. FAO. Global Forest Resources Assessment 2000. 2001.

51. Macintosh A. The Australia clause and REDD: a cautionary tale. Climatic Change. 2012;112: 169–188. doi:10.1007/s10584-011-0210-x

52. Marcos-Martinez R, Bryan BA, Connor JD, King D. Agricultural land-use dynamics: Assessing the relative importance of socioeconomic and biophysical drivers for more targeted policy. Land Use Policy. 2017;63: 53–66. doi:10.1016/j.landusepol.2017.01.011

53. Simmons BA, Marcos-Martinez R, Law EA, Bryan BA, Wilson KA. Frequent policy uncertainty can negate the benefits of forest conservation policy. Environmental Science & Policy. 2018;89: 401–411. doi:10.1016/j.envsci.2018.09.011

54. Rolfe J. Broadscale tree clearing in Queensland. Agenda. 2000;7: 219–236.

55. Simmons BA, Law EA, Marcos-Martinez R, Bryan BA, McAlpine C, Wilson KA. Spatial and temporal patterns of land clearing during policy change. Land Use Policy. 2018;75: 399–410. doi:10.1016/j.landusepol.2018.03.049

56. Kehoe J. Environmental law making in Queensland the Vegetation Management Act 1999 (Qld). Environmental and planning law journal. 2009;26: 392–410.

57. McGrath CJ. End of broadscale clearing in Queensland. Environment and Planning Law Journal. 2007;24: 5–13.

58. Australian Government. Australian National Greenhouse Accounts: Australian Land Use, Land Use Change and Forestry Emissions Projections to 2030. 2013.

59. Department of Environment and Resource Management. Analysis of Woody Vegetation Clearing Rates in Queensland: Supplementary report to Land cover change in Queensland 2008–09. 2010.

60. Garnaut R. Chapter 22: Transforming Rural Land Use. 2008.

61. Simmons BA, Wilson KA, Marcos-Martinez R, Bryan BA, Holland O, Law EA. Effectiveness of regulatory policy in curbing deforestation in a biodiversity hotspot. Environ Res Lett. 2018;13: 124003. doi:10.1088/1748-9326/aae7f9

62. Rhodes JR, Cattarino L, Seabrook L, Maron M. Assessing the effectiveness of regulation to protect threatened forests. Biological Conservation. Elsevier; 2017;216: 33–42. doi:10.1016/j.biocon.2017.09.020

63. Chiu G, Lockhart R, Routledge R. Bent-Cable Regression Theory and Applications. Journal of the American Statistical Association. 2006;101: 542–553. doi:10.1198/016214505000001177

64. Chiu GS, Lockhart RA. Bent-cable regression with autoregressive noise. Canadian Journal of Statistics. 2010;38: 386–407. doi:10.1002/cjs.10070

65. Khan SA, Chiu G, Dubin JA. Atmospheric Concentration of Chlorofluorocarbons: Addressing the Global Concern with the Longitudinal Bent- Cable Model. CHANCE. 2009;22: 8–17.

66. Toms JD, Lesperance ML. Piecewise regression: a tool for identifying ecological thresholds. Ecology. 2003;84: 2034–2041. doi:10.1890/02-0472

67. Khan SA, Rana M, Li L, Dubin JA. A Statistical Investigation to Monitor and Understand Atmospheric CFC Decline with the Spatial-longitudinal Bent-cable Model. International Journal of Statistics and Probability. 2012;1. doi:10.5539/ijsp.v1n2p56

68. Department of Agriculture, Fisheries and Forestry. State of Queensland agriculture report. State of Queensland; 2014.

69. Department of Science, Information Technology and Innovation. Land cover change in Queensland 2012–13 and 2013–14. 2015.

70. Australian Department of the Environment. Australian National Inventory System Forest Extent Data (version 12). Canberra; 2016.

71. Australian Department of the Environment. Human Induced Forest Extent & Change (version 11). Canberra; 2015.

72. Australian Greenhouse Office. Land clearing: a social history. National Carbon Accounting System Technical Report No. 4. Canberra, Australia; 2000.

73. Dwyer JM, Fensham RJ, Butler DW, Buckley YM. Carbon for conservation: Assessing the potential for win–win investment in an extensive Australian regrowth ecosystem. Agriculture, Ecosystems & Environment. 2009;134: 1–7. doi:10.1016/j.agee.2009.06.003

74. Fensham RJ, Guymer GP. Carbon accumulation through ecosystem recovery. Environmental Science & Policy. 2009;12: 367–372. doi:10.1016/j.envsci.2008.12.002

75. SEABROOK L, MCALPINE C, FENSHAM R. Cattle, crops and clearing: Regional drivers of landscape change in the Brigalow Belt, Queensland, Australia, 1840–2004. Landscape and Urban Planning. 2006;78: 373–385. doi:10.1016/j.landurbplan.2005.11.007

76. Reside AE, Beher J, Cosgrove AJ, Evans MC, Seabrook L, Silcock JL, et al. Ecological consequences of land clearing and policy reform in Queensland. Pacific Conservation Biology. 2017;23: 219–230. doi:10.1071/PC17001

77. McCulloch Robertson. New vegetation management laws pass [Internet]. 10 May 2018 [cited 21 Jun 2019]. Available: https://www.mccullough.com.au/2018/05/10/new-vegetation-management-laws-pass/

78. Furby S. Land cover change : specifications for remote sensing analysis. National Carbon Accounting System Technical Report No. 9. 2002.

79. Lehmann EA, Wallace JF, Caccetta P a., Furby SL, Zdunic K. Forest cover trends from time series Landsat data for the Australian continent. International Journal of Applied Earth Observation and Geoinformation. 2013;21: 453–462. doi:10.1016/j.jag.2012.06.005

80. ABARES. Land Use of Australia, Version 4, 2005/2006. 2010.

81. State of Queensland. Vegetation Management Act 1999 [Internet]. 1999. Available: http://www.austlii.edu.au/au/legis/qld/consol_act/vma1999212/

82. Hijmans RJ, van Etten J. raster: Geographic analysis and modeling with raster data [Internet]. R package version 2.0-12; 2012. Available: http://CRAN.R-project.org/package=raster

83. R Development Core Team. R: A language and environment for statistical computing [Internet]. Vienna, Austria: R Foundation for Statistical Computing; 2017. Available: http://www.R-project.org

84. The World Bank. World Development Indicators [Internet]. 24 Sep 2015 [cited 30 Sep 2015]. Available: http://data.worldbank.org/data-catalog/world-development-indicators

85. Eisner R, Seabrook LM, McAlpine CA. Are changes in global oil production influencing the rate of deforestation and biodiversity loss? Biological Conservation. 2016;196: 147–155. doi:10.1016/j.biocon.2016.02.017

86. U.S. Energy Information Administration. EIA Short-Term Energy Outlook, November 2015. 2015.

87. ABARES. Agricultural commodity statistics 2015. Canberra; 2015.

88. Rolfe J. Economics of vegetation clearing in Queensland. The Rangeland Journal. 2002;24: 152–169.

89. Jones DA, Wang W, Fawcett R. High-quality spatial climate data-sets for Australia. Australian Meteorological and Oceanographic Journal. 2009;58: 233.

90. Bureau of Meteorology. Average annual & monthly maximum, minimum, & mean temperature [Internet]. 2017. Report No.: BOM product code: IDCKAX1A20. Available: http://www.bom.gov.au/climate/austmaps/about-temp-maps.shtml

91. Bureau of Meteorology. Daily 9am vapour pressure for Australia [Internet]. 2017. Report No.: BOM product code: IDCKAVDA96. Available: http://www.bom.gov.au/climate/austmaps/about-vprp-maps.shtml

92. Bureau of Meteorology. Average annual, seasonal and monthly rainfall [Internet]. 2017. Report No.: BOM product code: IDCJCM004. Available: http://www.bom.gov.au/climate/austmaps/about-vprp-maps.shtml

93. Perry M. rasterstats 0.7.0 [Internet]. 2015. Available: https://github.com/perrygeo/python-rasterstats

94. Geoscience Australia. Shuttle Radar Topographic Mission (SRTM) Level 2 Elevation Data: 1 second Digital Surface Model (SRTM-DSM Level 2) v1.0 [Internet]. 2017. Available: http://data.aims.gov.au/metadataviewer/uuid/67b96045-2dae-47fb-90b9-4f6c5325764e

95. Pressey RL, Ferrier S, Hager TC, Woods CA, Tully SL, Weinman KM. How well protected are the forests of north-eastern New South Wales? -- Analyses of forest environments in relation to formal protection measures, land tenure, and vulnerability to clearing. Forest Ecology and Management. 1996;85: 311–333.

96. Angelsen A, Kaimowitz D. Rethinking the Causes of Deforestation: Lessons from Economic Models. World Bank Res Obs. 1999;14: 73–98. doi:10.1093/wbro/14.1.73

97. Tucker CJ, Pinzon JE, Brown ME. Global Inventory Modeling and Mapping Studies [Internet]. College Park, Maryland: Global Land Cover Facility, University of Maryland; 2014. Report No.: 3g.v0. Available: https://nex.nasa.gov/nex/projects/1349/

98. Geoscience Australia. Australian Land Tenure 1993. Canberra; 1993.

99. Earnest A, Morgan G, Mengersen K, Ryan L, Summerhayes R, Beard J. Evaluating the effect of neighbourhood weight matrices on smoothing properties of Conditional Autoregressive (CAR) models. International Journal of Health Geographics. 2007;6: 54. doi:10.1186/1476-072X-6-54

100. Sturtz S, Ligges U, Gelman A, others. R2WinBUGS: a package for running WinBUGS from R. Journal of Statistical software. 2005;12: 1–16.

101. Gelman A, Carlin JB, Stern HS, Dunson DB, Vehtari A, Rubin DB. Bayesian Data Analysis, Third Edition. CRC Press; 2013.

102. Kery M. Introduction to WinBUGS for Ecologists: Bayesian Approach to Regression, ANOVA, Mixed Models and Related Analyses. Academic Press; 2010.

103. Lunn DJ, Thomas A, Best N, Spiegelhalter D. WinBUGS - A Bayesian modelling framework: Concepts, structure, and extensibility. Statistics and Computing. 2000;10: 325–337. doi:10.1023/A:1008929526011

104. 3218.0 – Regional Population Growth, Australia, 2014–15. Canberra: Australian Bureau of Statistics; 2016 Mar.

105. State of Queensland. List of vegetation clearing exemptions [Internet]. Brisbane: Department of Natural Resources and Mines; 2015. Available: https://www.dnrm.qld.gov.au/¨data/assets/pdf file/0009/847800/vegetation-clearing-exemptions.pdf

106. Rhodes JR, Wiegand T, McAlpine CA, Callaghan J, Lunney D, Bowen M, et al. Modeling species' distributions to improve conservation in semiurban landscapes: Koala case study. Conservation Biology. 2006;20: 449–459. doi:10.1111/j.1523-1739.2006.00330.x

107. Rhodes J, Beyer H, Preece H, McAlpine C. South East Queensland Koala Population Modelling Study. Brisbane, Australia: UniQuest; 2015.

108. Nelder VJ, Niehus RE, Wilson BA, McDonald WJF, Ford AJ, Accad A. The Vegetation of Queensland. Descriptions of Broad Vegetation Groups. Brisbane: Queensland Herbarium, Department of Science, Information Technology and Innovation; 2017. Report No.: Version 3.0.

109. Gunders J. Weed and feed: Graziers push for mulga reclassification as woody weed and stock feed. ABC Rural. 2017. Available: http://www.abc.net.au/news/rural/2017-06-19/graziers-want-mulga-reclassified-as-woody-weed-and-drought-feed/8624052. Accessed 29 Jun 2017.

110. Pressey RL, Ferrier S, Hager TC, Woods CA, Tully SL, Weinman KM. How well protected are the forests of north-eastern New South Wales? -- Analyses of forest environments in relation to formal protection measures, land tenure, and vulnerability to clearing. Forest Ecology and Management. 1996;85: 311–333.

111. Commonwealth of Australia. Australian Land Use, Land Use-Change and Forestry emissions projections. Canberra; 2015 Mar.

112. Marinoni O, Navarro Garcia J, Marvanek S, Prestwidge D, Clifford D, Laredo LA. Development of a system to produce maps of agricultural profit on a continental scale: An example for Australia. Agricultural Systems. 2012;105: 33–45.

113. Bartel R, Barclay E. Motivational Postures and Compliance with Environmental Law in Australian Agriculture. Journal of Rural Studies. 2011;27: 153–170.

114. Bulinski J, Enright R, Tomsett N. Tree clearing in Australia: Its Contribution to Climate Change. 2016 Feb.

115. Commonwealth of Australia. Quarterly Update of Australia's National Greenhouse Gas Inventory: September 2015. Canberra: Commonwealth of Australia; 2016 Mar.

116. Rue H, Martino S, Chopin N. Approximate Bayesian inference for latent Gaussian models using inte- grated nested Laplace approximations (with discussion). Journal of the Royal Statistical Society, Series B. 2009;71: 319–392.

# Appendix

## 1.1 Additional details on regulatory policies for controlling deforestation in Queensland, 1990-2014

Table A1.1: Policy changes corresponding to Figure 2

|     | Year | Policy name | Details |
| --- | --- | --- | --- |
| (a) | 1995 | *Land Act 1994* | Introduced to control native vegetation clearing on leasehold and State lands. Clearing on freehold still regulated by local governments under the *Local Government Act (LGA) 1993* and the *Planning and Environment Act 1990* |
| (b) | 2000 | *Vegetation Management Act 1999* (VMA) | Enabled state-level regulation of native vegetation clearing on freehold land. Clearing of regrowth vegetation still allowed. |
| (c) | 2003 |  | A moratorium on tree clearing applications imposed by the Queensland Government in May 2003 |
| (d) | 2004 | *Vegetation Management and Other Legislation Bill 2004* | Regulation of vegetation clearing on leasehold and State land removed from the *Land Act 1994* and placed under the VMA. A ballot for clearing permits totalling 500,000 hectares was held in September 2004. Provided $150 million of financial assistance over five years |
| (e) | 2006 |  | All clearing permits issued under the ballot held in 2004 expired on 31 December 2006. |
| (f) | 2009 | *Vegetation Management and Other Legislation Amendment Act 2009* | VMA amended to protect 'high value regrowth' regrowth (vegetation not cleared since 31 December 1989) |
| (g) | 2012 |  | Review ordered into the enforcement of the VMA; all investigations into noncompliance suspended |
| (h) | 2013 | Vegetation Management Regulation 2012 under the *Vegetation Management Framework Amendment Act 2013* | Introduced a series of self-assessable codes for vegetation clearing, removed regulations on 'high value' regrowth clearing, introduced permitted clearing for necessary environmental clearing, high and irrigated high value agricultural clearing |

## 1.2 R code and output for covariate selection process

```r
#final df
newdf<-read.csv("newdf4_new_Evans13June_qld.csv")
newdf <- data.frame(newdf[ ,-1])
newdf_ayy <- data.frame(newdf[ ,c(1,2,6)])

#qld lgas
qldlgas<-read.csv("qlflgasforanalysis_13June.csv")

########Climate covariates

lgaclimate<-read.csv("lgaClimate.csv")
lgadem<-read.csv("lga_dem_ndvi.csv")

#select out the LGAs we're analysing
qldlgaclimate<-merge(lgaclimate, qldlgas, by=intersect(names(lgaclimate), names(qldlgas)))
qldlgadem<-merge(lgadem, qldlgas, by.x="LGA_CODE11", by.y="LGA_CODE")

#add ayy (response variable y[it])
ayyqldlgaclimate<-merge(qldlgaclimate, newdf_ayy, by=intersect(names(lgaclimate), names(newdf_ayy)))
ayyqldlgadem<-merge(qldlgadem, newdf_ayy, by.x="LGA_CODE11", by.y="LGA_CODE")

#Log transform climate variables
ayyqldlgaclimate$logAnnualMinMonthRain<-log10(ayyqldlgaclimate$AnnualMinMonthRain+1)
ayyqldlgaclimate$logAnnualMaxMonthRain<-log10(ayyqldlgaclimate$AnnualMaxMonthRain+1)
ayyqldlgaclimate$logAnnualRain<-log10(ayyqldlgaclimate$AnnualRain+1)
ayyqldlgaclimate$logVap<-log10(ayyqldlgaclimate$AnnualMeanVpd09)

####Scatterplots
## Untransformed rainfall and vapour pressure
plot(qldlgaclimate[, c(2:5)])
```

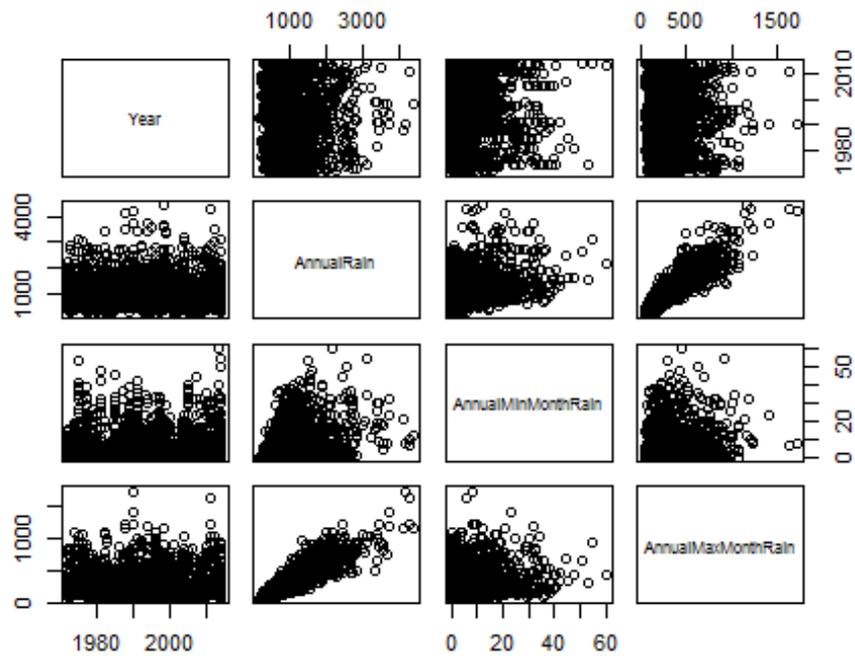

```r
cor(qldlgaclimate[, c(2:5)])
```

```
##                           Year     AnnualRain AnnualMinMonthRain
## Year                1.000000000 -0.004828387         0.09340903
## AnnualRain         -0.004828387  1.000000000         0.36668961
## AnnualMinMonthRain  0.093409030  0.366689608         1.00000000
## AnnualMaxMonthRain -0.041190610  0.889617485         0.18610639
##                    AnnualMaxMonthRain
## Year                      -0.04119061
## AnnualRain                 0.88961748
## AnnualMinMonthRain         0.18610639
## AnnualMaxMonthRain         1.00000000
```

```r
#ayy, year and log transformed rainfall and vapour pressure
plot(ayyqldlgaclimate[, c(2,13:17)])
```

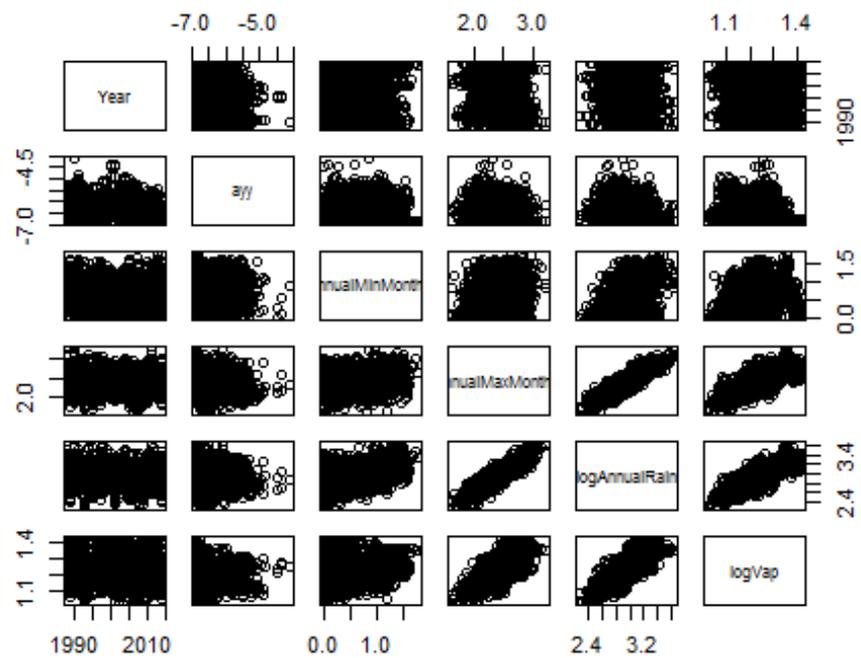

**#All rain and vapour variables pertain to moisture, and none stands out**
**#as an obvious choice. WE CHOOSE VAPOUR PRESSURE.**

##Temperature
## Untransformed temperature
plot(ayyqldlgaclimate[, c(2,13, 7:10)])

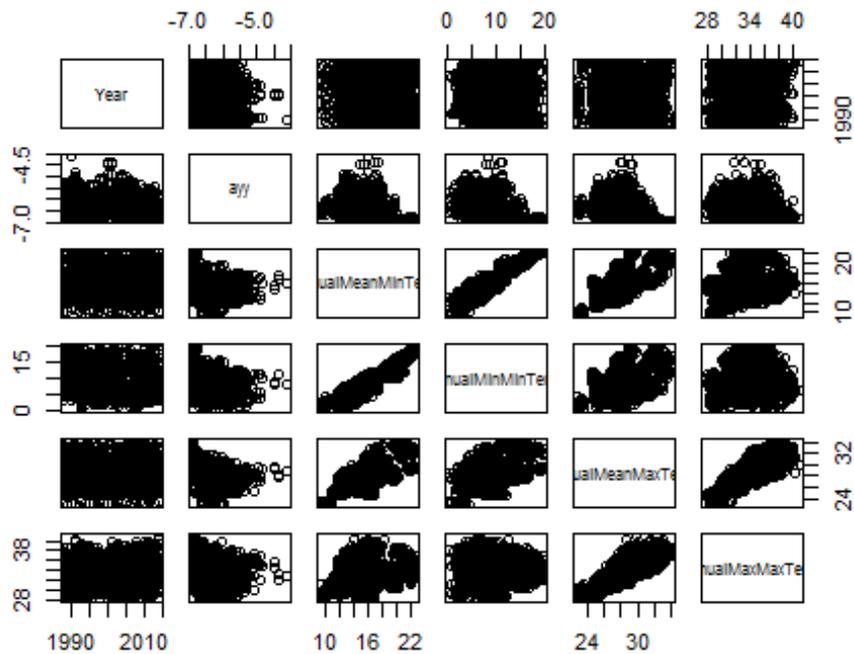

```r
cor(ayyqldlgaclimate[, c(2,13, 7:10)])
```

```
##                         Year        ayy AnnualMeanMinTemp
## Year              1.00000000 -0.1841749       -0.02188517
## ayy              -0.18417486  1.0000000       -0.40106774
## AnnualMeanMinTemp -0.02188517 -0.4010677        1.00000000
## AnnualMinMinTemp  -0.04240353 -0.3730308        0.94286401
## AnnualMeanMaxTemp  0.02034789 -0.3887173        0.70818730
## AnnualMaxMaxTemp   0.07129730 -0.2006276        0.28036365
##                   AnnualMinMinTemp AnnualMeanMaxTemp AnnualMaxMaxTemp
## Year                   -0.04240353        0.02034789       0.07129730
## ayy                    -0.37303079       -0.38871727      -0.20062755
## AnnualMeanMinTemp       0.94286401        0.70818730       0.28036365
## AnnualMinMinTemp        1.00000000        0.53479279       0.07170248
## AnnualMeanMaxTemp       0.53479279        1.00000000       0.82098644
## AnnualMaxMaxTemp        0.07170248        0.82098644       1.00000000
```

```r
#Log transform
ayyqldlgaclimate$logAnnualMeanMinTemp<-log10(ayyqldlgaclimate$AnnualMeanMinTemp+1)
ayyqldlgaclimate$logAnnualMinMinTemp<-log10(ayyqldlgaclimate$AnnualMinMinTemp+1)
ayyqldlgaclimate$logAnnualMeanMaxTemp <-log10(ayyqldlgaclimate$AnnualMeanMaxTemp+1
)
ayyqldlgaclimate$logAnnualMaxMaxTemp<-log10(ayyqldlgaclimate$AnnualMaxMaxTemp+1)

##ayy, year and log transformed temperature
plot(ayyqldlgaclimate[, c(2,13, 18:21)])
```

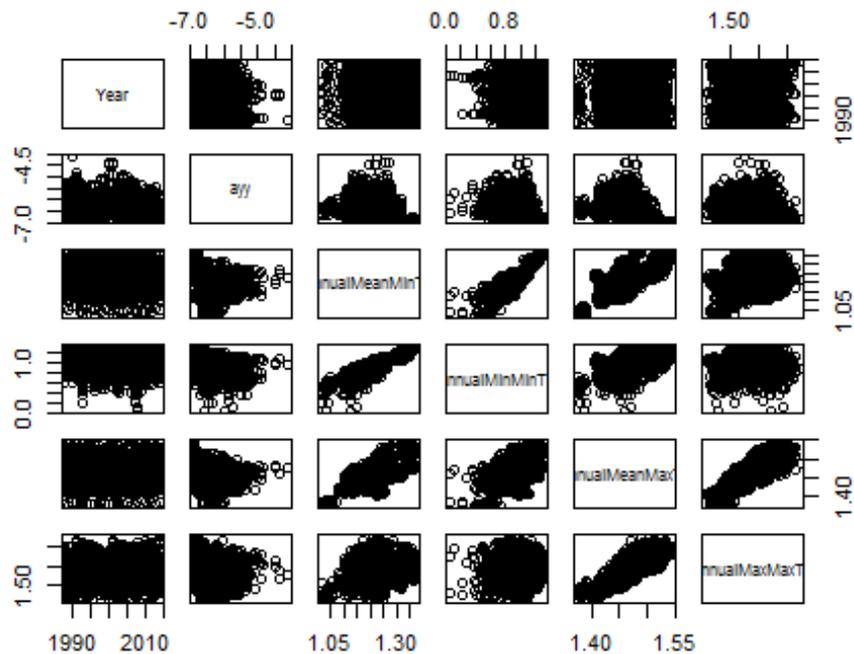

```r
cor(ayyqldlgaclimate[, c(2,13, 18:21)])
```

```
##                          Year        ayy logAnnualMeanMinTemp
## Year               1.00000000 -0.1841749          -0.02394022
## ayy               -0.18417486  1.0000000          -0.36793368
## logAnnualMeanMinTemp -0.02394022 -0.3679337          1.00000000
## logAnnualMinMinTemp  -0.05636583 -0.3033580          0.90757342
## logAnnualMeanMaxTemp  0.02121855 -0.3730941          0.70875948
## logAnnualMaxMaxTemp   0.07112957 -0.1945425          0.29200122
##                      logAnnualMinMinTemp logAnnualMeanMaxTemp
## Year                         -0.05636583           0.02121855
## ayy                          -0.30335801          -0.37309411
## logAnnualMeanMinTemp          0.90757342           0.70875948
## logAnnualMinMinTemp           1.00000000           0.49771370
## logAnnualMeanMaxTemp          0.49771370           1.00000000
## logAnnualMaxMaxTemp           0.04923780           0.83031241
##                      logAnnualMaxMaxTemp
## Year                          0.07112957
## ayy                          -0.19454247
## logAnnualMeanMinTemp          0.29200122
## logAnnualMinMinTemp           0.04923780
## logAnnualMeanMaxTemp          0.83031241
## logAnnualMaxMaxTemp           1.00000000
```

#None of temp variables stands out as an obvious choice.

#WE CHOOSE MEANMAXTEMP (highest corr with response ayy).

#########Biophysical covariates

##Untransformed elevation
```r
plot(ayyqldlgadem[,c(21,5:10)])
```

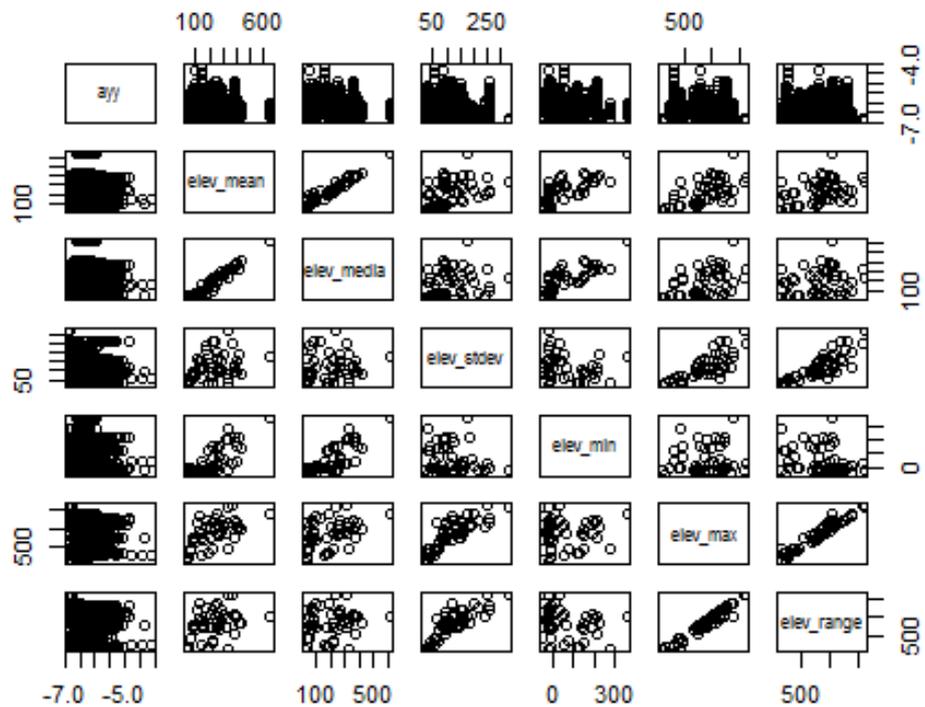

```r
cor(ayyqldlgadem[,c(21,5:10)])
```

```
##                    ayy    elev_mean  elev_media   elev_stdev     elev_min
## ayy         1.00000000   0.02855343  0.05356424  -0.07107317   0.07944516
## elev_mean   0.02855343   1.00000000  0.96746196   0.35055171   0.77298668
## elev_media  0.05356424   0.96746196  1.00000000   0.12652407   0.82481262
## elev_stdev -0.07107317   0.35055171  0.12652407   1.00000000  -0.16233502
## elev_min    0.07944516   0.77298668  0.82481262  -0.16233502   1.00000000
## elev_max    0.07153512   0.59647558  0.44577600   0.80609237   0.11493058
## elev_range  0.05087606   0.39582254  0.23235147   0.84490271  -0.14230368
##              elev_max   elev_range
## ayy         0.07153512   0.05087606
## elev_mean   0.59647558   0.39582254
## elev_media  0.44577600   0.23235147
## elev_stdev  0.80609237   0.84490271
## elev_min    0.11493058  -0.14230368
## elev_max    1.00000000   0.96690896
## elev_range  0.96690896   1.00000000
```

```r
#Untransformed slope
plot(ayyqldlgadem[,c(21,11:16)])
```

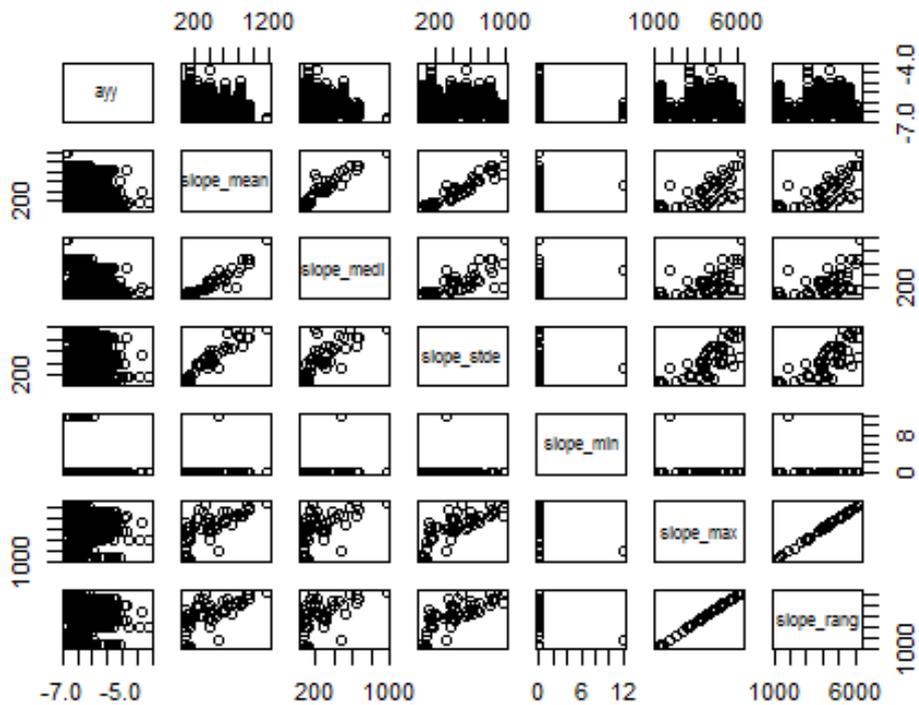

```r
cor(ayyqldlgadem[,c(21,11:16)])
```

```
##                       ayy   slope_mean  slope_medi   slope_stde   slope_min
## ayy           1.000000000  -0.03058673 -0.04887301 -0.001940041 -0.07684461
## slope_mean   -0.030586728   1.00000000  0.92959860  0.949089978  0.05870802
## slope_medi   -0.048873013   0.92959860  1.00000000  0.776482499  0.15586982
## slope_stde   -0.001940041   0.94908998  0.77648250  1.000000000 -0.05860837
## slope_min    -0.076844608   0.05870802  0.15586982 -0.058608371  1.00000000
## slope_max     0.034588532   0.66269723  0.52211208  0.772201198 -0.23192687
## slope_rang    0.034672973   0.66243607  0.52177180  0.772052776 -0.23308799
##              slope_max  slope_rang
## ayy         0.03458853  0.03467297
## slope_mean  0.66269723  0.66243607
## slope_medi  0.52211208  0.52177180
## slope_stde  0.77220120  0.77205278
## slope_min  -0.23192687 -0.23308799
## slope_max   1.00000000  0.99999929
## slope_rang  0.99999929  1.00000000
```

```r
##All elevation variables are highly correlated with each other,

##and same with slope variables, so select mean and range for now

#Log transform slope and elev mean and range
ayyqldlgadem$logelev_mean<-log10(ayyqldlgadem$elev_mean+1)
ayyqldlgadem$logelev_range<-log10(ayyqldlgadem$elev_range+1)
ayyqldlgadem$logslope_mean <-log10(ayyqldlgadem$slope_mean+1)
ayyqldlgadem$logslope_rang<-log10(ayyqldlgadem$slope_rang+1)

plot(ayyqldlgadem[,c(21:25)])
```

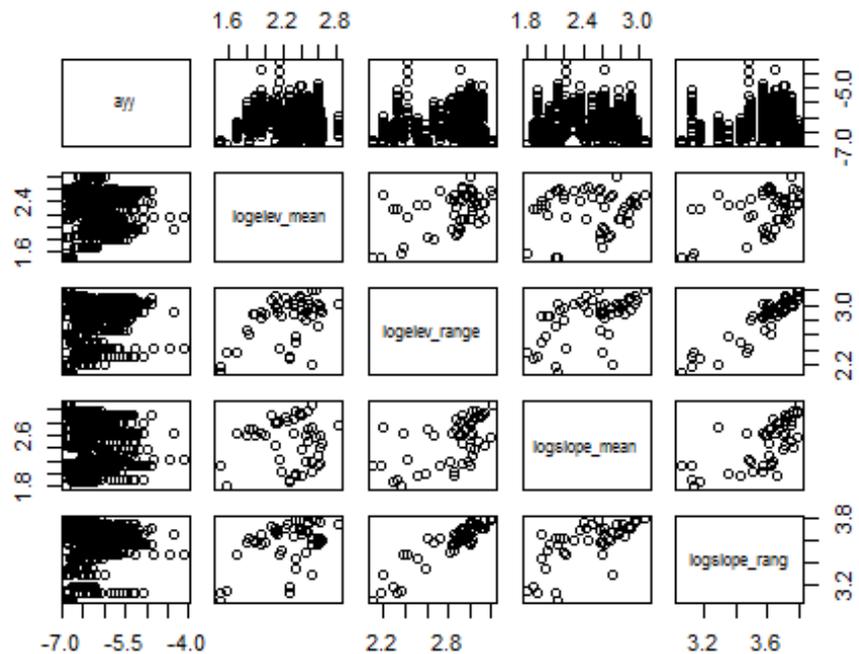

```r
cor(ayyqldlgadem[,c(21:25)])
```

```
##                      ayy logelev_mean logelev_range logslope_mean
## ayy           1.00000000   0.09355166     0.1390674    0.06358031
## logelev_mean  0.09355166   1.00000000     0.5245611    0.07242372
## logelev_range 0.13906743   0.52456115     1.0000000    0.54859437
## logslope_mean 0.06358031   0.07242372     0.5485944    1.00000000
## logslope_rang 0.11090982   0.48167175     0.9197467    0.67160320
##               logslope_rang
## ayy               0.1109098
## logelev_mean      0.4816717
## logelev_range     0.9197467
## logslope_mean     0.6716032
## logslope_rang     1.0000000
```

```r
##None stands out as obvious choice, so
##CHOOSE JUST ELEV_MEAN for interpretability

##NDVI
#look at ayy vs ndvi
plot(ayyqldlgadem[, c(21, 17:19)])
```

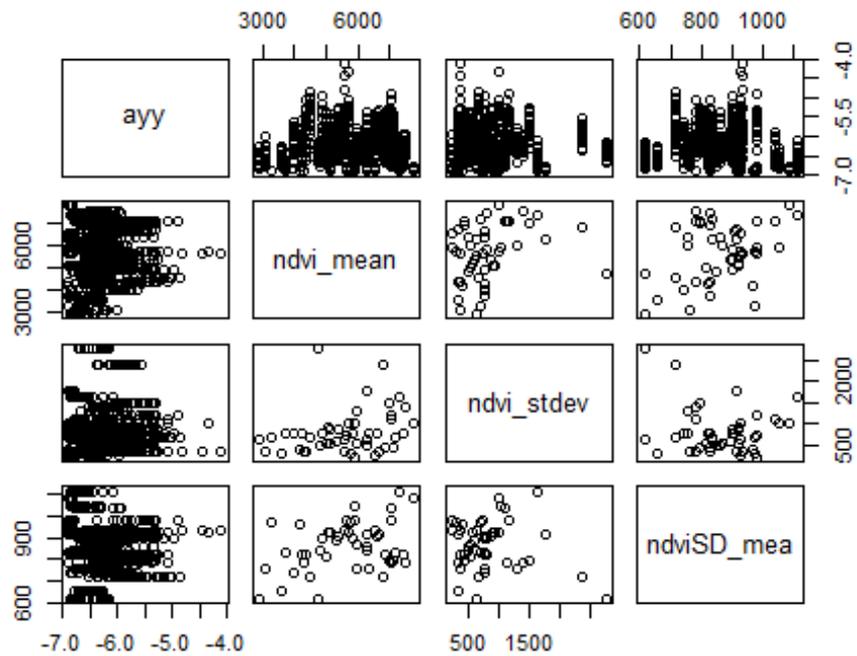

```r
cor(ayyqldlgadem[, c(21, 17:19)])

##                     ayy  ndvi_mean   ndvi_stdev  ndviSD_mea
## ayy         1.00000000  0.1098183   0.06964855  -0.1898965
## ndvi_mean   0.10981828  1.0000000   0.27468511   0.3399635
## ndvi_stdev  0.06964855  0.2746851   1.00000000  -0.1384215
## ndviSD_mea -0.18989645  0.3399635  -0.13842150   1.0000000

#Log transform
ayyqldlgadem$logndvi_mean<-log10(ayyqldlgadem$ndvi_mean+1)
ayyqldlgadem$logndvi_stdev <-log10(ayyqldlgadem$ndvi_stdev+1)
ayyqldlgadem$logndviSD_mea<-log10(ayyqldlgadem$ndviSD_mea+1)

##ayy, year and log transformed NDVI
plot(ayyqldlgadem[, c(21, 26:28)])
```

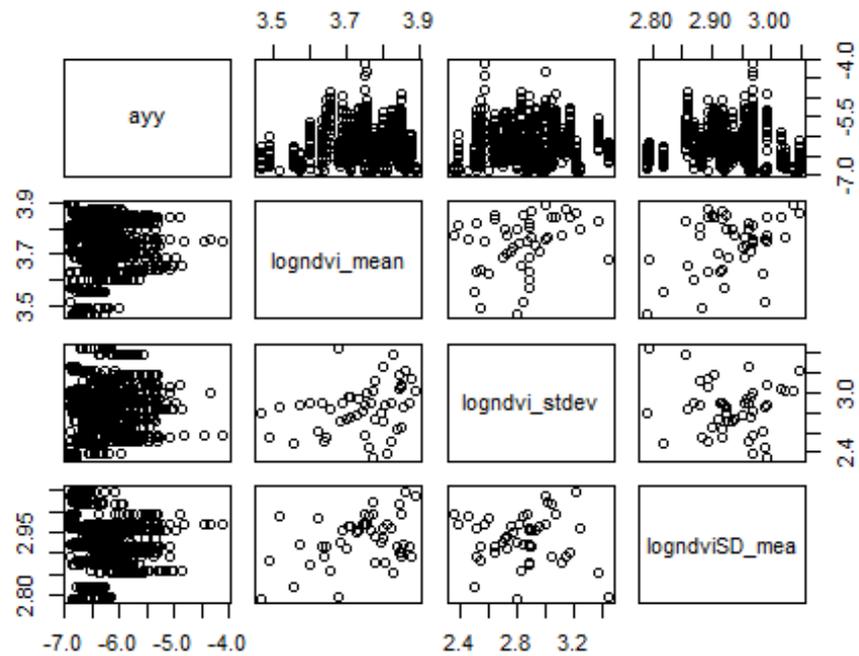

```
cor(ayyqldlgadem[, c(21, 26:28)])

##                         ayy logndvi_mean logndvi_stdev logndviSD_mea
## ayy            1.0000000    0.1363419     0.1059620    -0.1691899
## logndvi_mean   0.1363419    1.0000000     0.2923990     0.3763235
## logndvi_stdev  0.1059620    0.2923990     1.0000000    -0.0794819
## logndviSD_mea -0.1691899    0.3763235    -0.0794819     1.0000000

##CHOOSE NDIV_MEAN for interpretability

#Examine selected climate and biophysical variables together
select_dem<-ayyqldlgadem[, c(1,21,22,26)]
select_clim<-ayyqldlgaclimate[, c(1,2,13,17,20)]

plot(select_dem[-1])
```

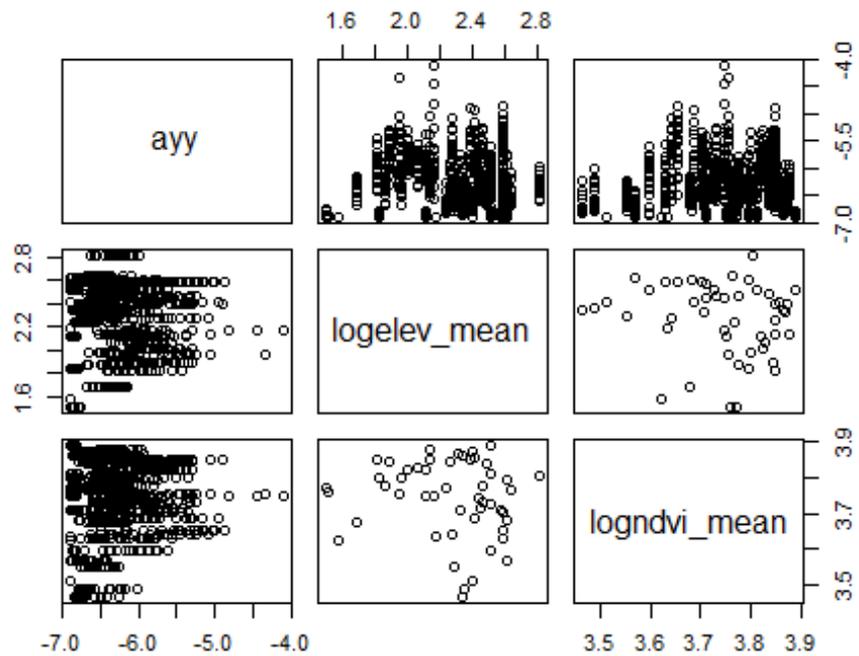

```r
cor(select_dem[-1])
```

```
##                     ayy logelev_mean logndvi_mean
## ayy          1.00000000   0.09355166    0.1363419
## logelev_mean 0.09355166   1.00000000   -0.1421020
## logndvi_mean 0.13634188  -0.14210203    1.0000000
```

```r
plot(select_clim[ ,2:5])
```

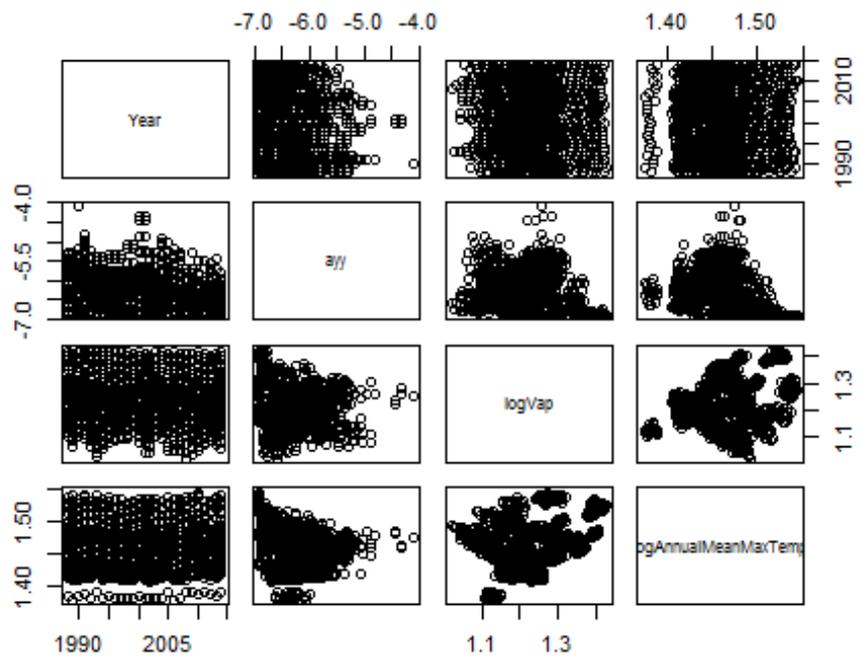

```r
cor(select_clim[ ,2:5])

##                              Year        ayy      logVap
## Year                  1.00000000 -0.1841749 -0.03429284
## ayy                  -0.18417486  1.0000000 -0.29181861
## logVap               -0.03429284 -0.2918186  1.00000000
## logAnnualMeanMaxTemp  0.02121855 -0.3730941  0.29744465
##                      logAnnualMeanMaxTemp
## Year                            0.02121855
## ayy                            -0.37309411
## logVap                          0.29744465
## logAnnualMeanMaxTemp            1.00000000

#merge
select_newdf<-merge(select_dem[-2], select_clim, by.x="LGA_CODE11", by.y="LGA_CODE
")
```

### Land tenure

```r
tenure<-read.csv("proptenure_tencov_14-6-17_qldonly.csv")
tenure<-tenure[,c(2,6)]
select_newdf<-merge(select_newdf, tenure, by.x="LGA_CODE11", by.y="LGA_CODE")

#Plot with land tenure
plot(select_newdf[,c(5,2,3,6,7,8)])
```

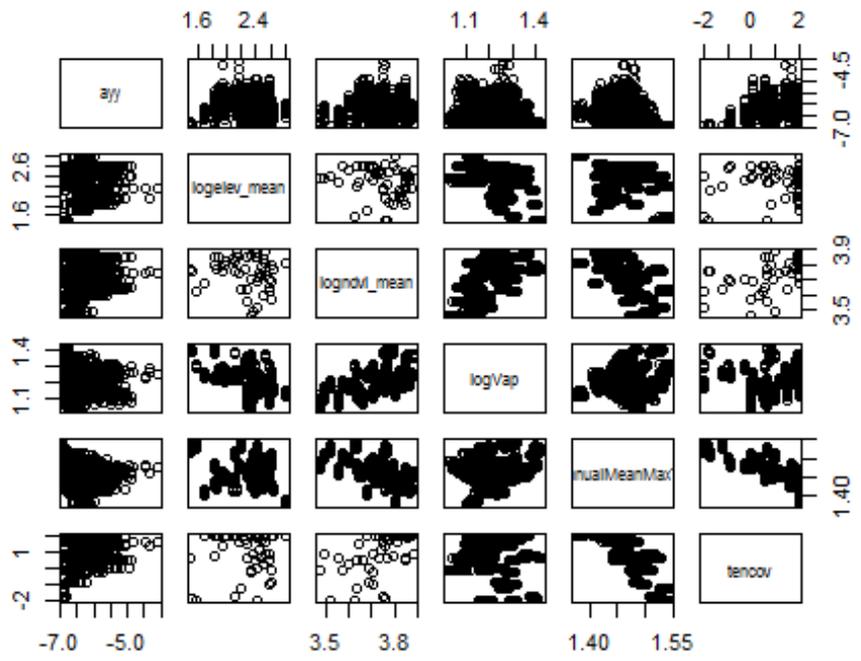

```r
cor(select_newdf[,c(5,2,3,6,7,8)])
```

```
##                          ayy logelev_mean logndvi_mean      logVap
## ayy               1.00000000   0.09355166    0.1363419 -0.2918186
## logelev_mean      0.09355166   1.00000000   -0.1421020 -0.5563906
## logndvi_mean      0.13634188  -0.14210203    1.0000000  0.4686681
## logVap           -0.29181861  -0.55639061    0.4686681  1.0000000
## logAnnualMeanMaxTemp -0.37309411 -0.20994323  -0.6007638  0.2974447
## tencov            0.40014492   0.09059084    0.5522605 -0.2487662
##                  logAnnualMeanMaxTemp     tencov
## ayy                        -0.3730941  0.40014492
## logelev_mean               -0.2099432  0.09059084
## logndvi_mean               -0.6007638  0.55226053
## logVap                      0.2974447 -0.24876616
## logAnnualMeanMaxTemp        1.0000000 -0.84803929
## tencov                     -0.8480393  1.00000000
```

**##with VAPOUR in the mix, NDVI doesn't provide additional information**

##Macroeconomic covariates

```r
macrodf<-read.csv("macro_df_march2016.csv")
```

##Broad macro variables
```r
plot(macrodf[,c(1,3,5,6,17,19,20,21)])
```

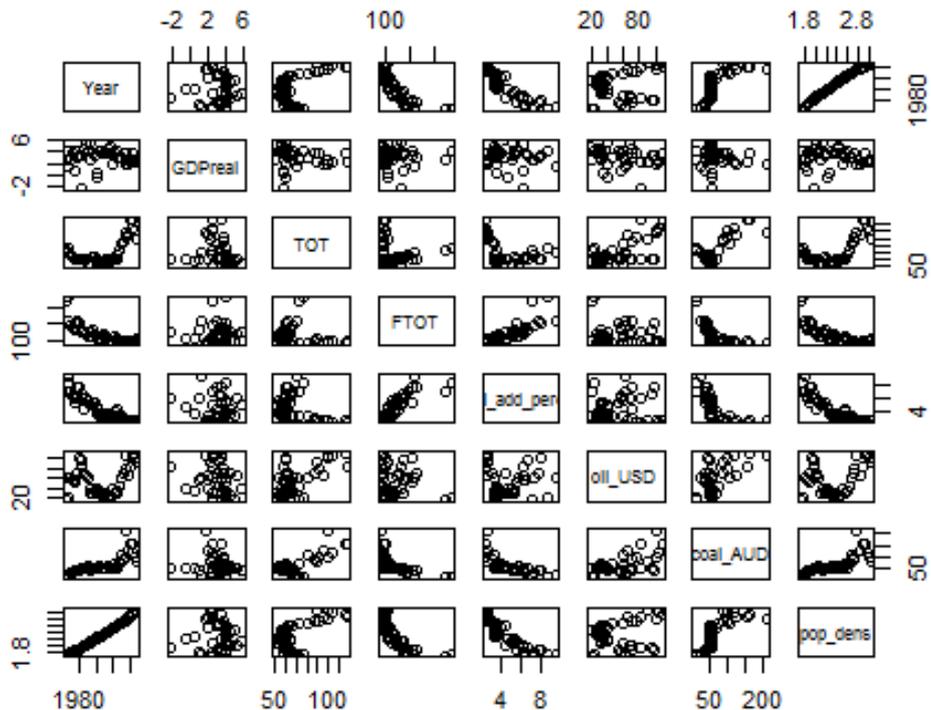

```
cor(macrodf[,c(1,3,5,6,17,19,20,21)])
```

```
##                           Year       GDPreal         TOT        FTOT
## Year               1.00000000   0.022366065   0.6231801  -0.80163985
## GDPreal            0.02236607   1.000000000  -0.1514287   0.02221356
## TOT                0.62318014  -0.151428718   1.0000000  -0.23000222
## FTOT              -0.80163985   0.022213560  -0.2300022   1.00000000
## Ag_val_add_percGDP -0.90649169 -0.079940970  -0.4294145   0.83579318
## oil_USD            0.27029412  -0.268715619   0.6456314  -0.16604924
## coal_AUD           0.76181960  -0.135916955   0.7863692  -0.61394550
## pop_dens           0.99617527  -0.004184176   0.6686562  -0.77176732
##                    Ag_val_add_percGDP    oil_USD    coal_AUD     pop_dens
## Year                      -0.90649169   0.2702941   0.7618196   0.996175267
## GDPreal                   -0.07994097  -0.2687156  -0.1359170  -0.004184176
## TOT                       -0.42941453   0.6456314   0.7863692   0.668656235
## FTOT                       0.83579318  -0.1660492  -0.6139455  -0.771767320
## Ag_val_add_percGDP         1.00000000  -0.0768725  -0.6750377  -0.890013900
## oil_USD                   -0.07687250   1.0000000   0.5844126   0.312729525
## coal_AUD                  -0.67503773   0.5844126   1.0000000   0.770744102
## pop_dens                  -0.89001390   0.3127295   0.7707441   1.000000000
```

**#CHOOSE GDPREAL, OILUSD, % AG to GDP TO MINIMIZE CONFOUNDING WITH YEAR**

**##Farm variables**
```
plot(macrodf[,c(1,8,10,11)])
```

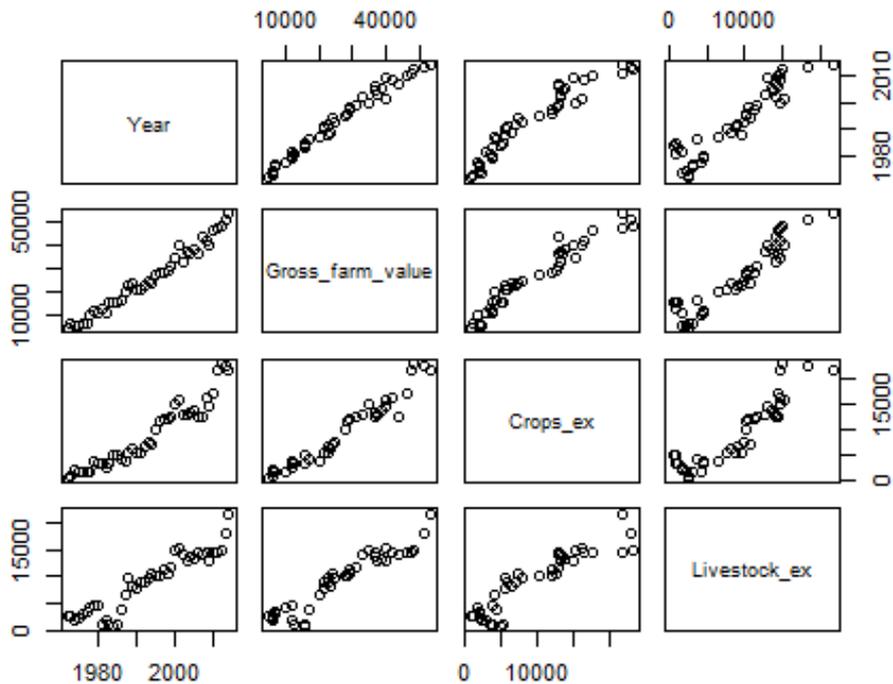

```r
cor(macrodf[,c(1,8,10,11)])
```

```
##                      Year Gross_farm_value  Crops_ex Livestock_ex
## Year            1.0000000        0.9891603 0.9566678    0.9335221
## Gross_farm_value 0.9891603        1.0000000 0.9705191    0.9436573
## Crops_ex        0.9566678        0.9705191 1.0000000    0.9146534
## Livestock_ex    0.9335221        0.9436573 0.9146534    1.0000000
```

```r
#Check livestock exports
select_macro<-macrodf[,c(1,17,19,20,11)]

##add ayy
select_macro_ayy<-merge(select_macro, newdf_ayy , by=intersect(names(select_macro)
, names(newdf_ayy)))

plot(select_macro_ayy[-6])
```

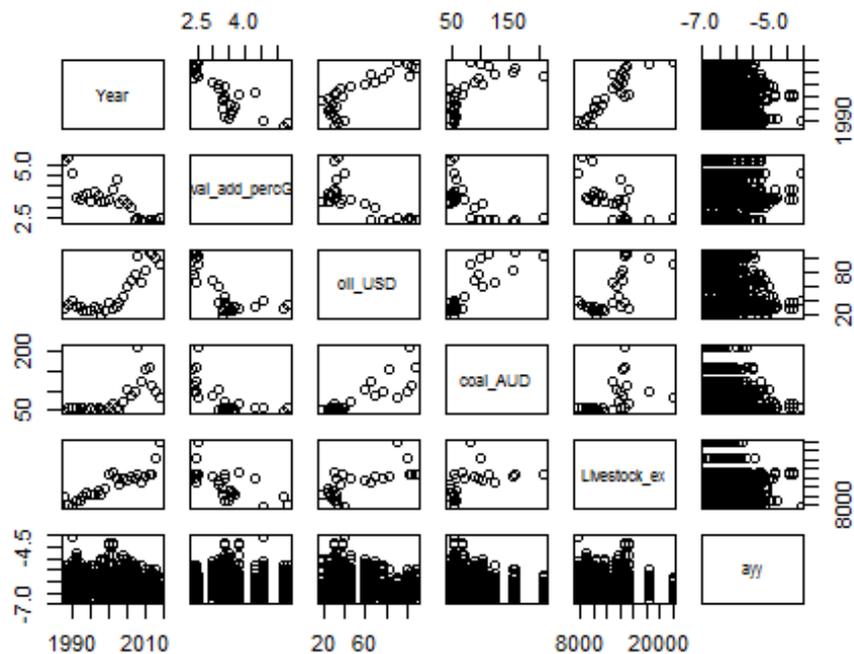

```r
cor(select_macro_ayy[-6])
```

```
##                       Year Ag_val_add_percGDP    oil_USD   coal_AUD
## Year             1.0000000         -0.8293058  0.8369721  0.6962969
## Ag_val_add_percGDP -0.8293058       1.0000000 -0.7003019 -0.6440887
## oil_USD          0.8369721         -0.7003019  1.0000000  0.8435394
## coal_AUD         0.6962969         -0.6440887  0.8435394  1.0000000
## Livestock_ex     0.8820347         -0.6339391  0.6918659  0.4672437
## ayy             -0.1841749          0.1624686 -0.1601019 -0.1589144
##                   Livestock_ex        ayy
## Year                 0.8820347 -0.1841749
## Ag_val_add_percGDP  -0.6339391  0.1624686
## oil_USD              0.6918659 -0.1601019
## coal_AUD             0.4672437 -0.1589144
## Livestock_ex         1.0000000 -0.1468161
## ayy                 -0.1468161  1.0000000
```

```r
select_macro_ayy$logAg<-log10(select_macro_ayy$Ag_val_add_percGDP)
select_macro_ayy$logoil<-log10(select_macro_ayy$oil_USD)
select_macro_ayy$loglive<-log10(select_macro_ayy$Livestock_ex)

plot(select_macro_ayy[c(1,7:10)])
```

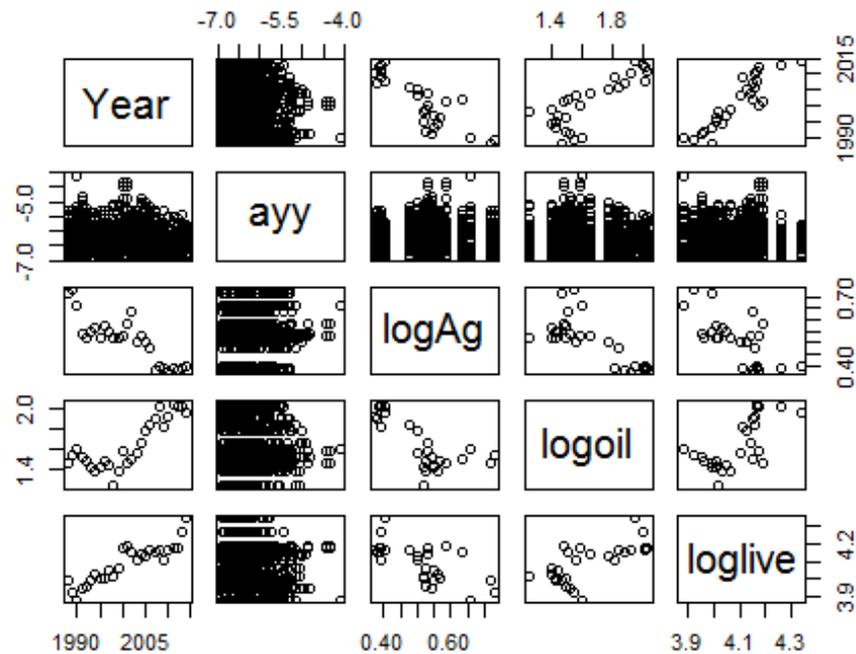

```r
cor(select_macro_ayy[c(1,7:10)])

##                Year        ayy       logAg     logoil    loglive
## Year      1.0000000 -0.1841749 -0.8596448  0.8230612  0.9061581
## ayy      -0.1841749  1.0000000  0.1740008 -0.1386798 -0.1520259
## logAg    -0.8596448  0.1740008  1.0000000 -0.7395789 -0.6865213
## logoil    0.8230612 -0.1386798 -0.7395789  1.0000000  0.6733363
## loglive   0.9061581 -0.1520259 -0.6865213  0.6733363  1.0000000
```

**##remove coal (too similar to oil) and livestock (too similar to year)**

```r
select_macro_ayy_<-select_macro_ayy[ ,c(-6,-7)]
```

########Combine final covariates

```r
select_macro$logAg<-log10(select_macro$Ag_val_add_percGDP)
select_macro$logoil<-log10(select_macro$oil_USD)
select_macro$loglive<-log10(select_macro$Livestock_ex)

finaldf_ayy<-merge(select_macro, select_newdf,
                   by=intersect(names(select_macro), names(select_newdf)))

finaldf_ayy<-finaldf_ayy[ ,c(-2,-3,-4,-5)]
```

*#Macro*
```r
plot(finaldf_ayy[,c(1,8,2,3)])
```

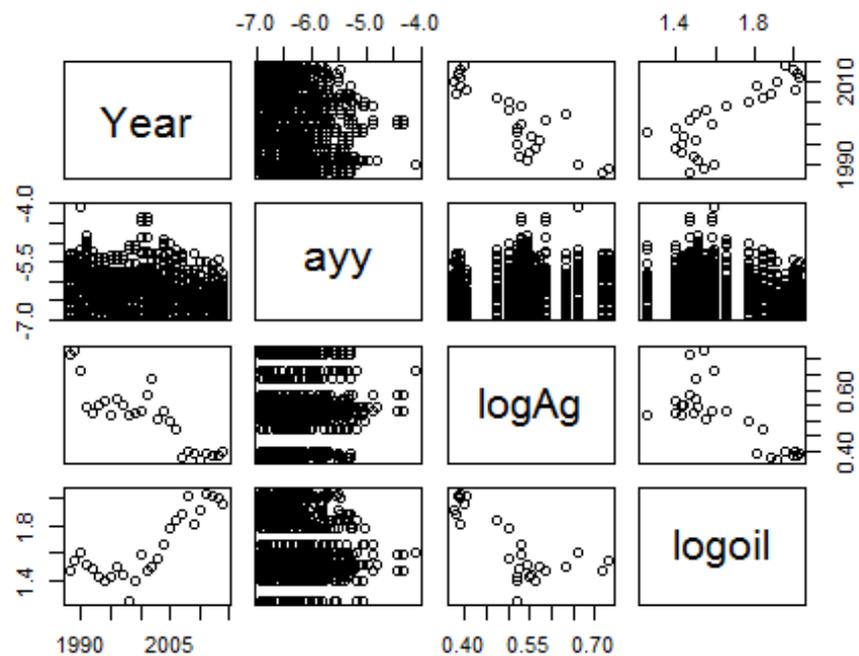

```
cor(finaldf_ayy[,c(1,8,2,3)])

##              Year        ayy      logAg     logoil
## Year    1.0000000 -0.1841749 -0.8596448  0.8230612
## ayy    -0.1841749  1.0000000  0.1740008 -0.1386798
## logAg  -0.8596448  0.1740008  1.0000000 -0.7395789
## logoil  0.8230612 -0.1386798 -0.7395789  1.0000000
```

```
#Geographical (biophysical and institutional)
plot(finaldf_ayy[,c(8,6,11)])
```

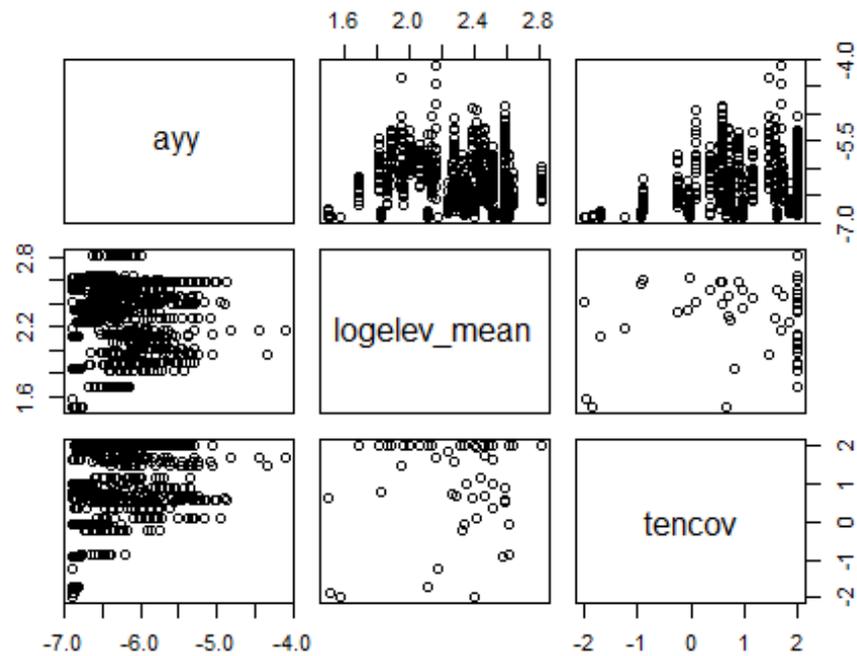

```
cor(finaldf_ayy[,c(1,8,6,11)])

##                        Year         ayy logelev_mean        tencov
## Year          1.000000e+00 -0.18417486   0.00000000 -3.449718e-21
## ayy          -1.841749e-01  1.00000000   0.09355166  4.001449e-01
## logelev_mean  0.000000e+00  0.09355166   1.00000000  9.059084e-02
## tencov       -3.449718e-21  0.40014492   0.09059084  1.000000e+00
```

```
#Climate
plot(finaldf_ayy[,c(8,9,10)])
```

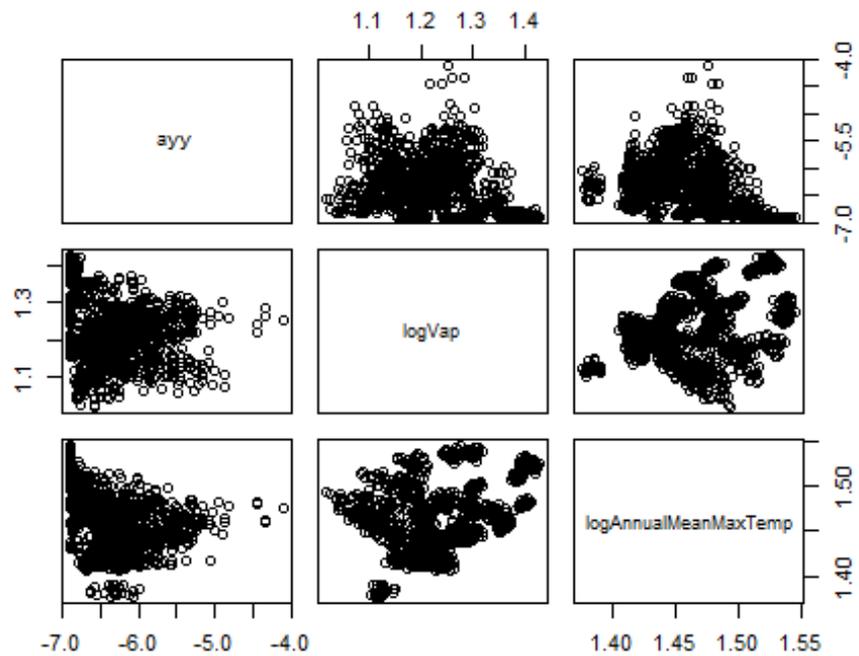

```
cor(finaldf_ayy[,c(1,8,9,10)])
```

```
##                              Year        ayy      logVap
## Year               1.00000000 -0.1841749 -0.03429284
## ayy               -0.18417486  1.0000000 -0.29181861
## logVap            -0.03429284 -0.2918186  1.00000000
## logAnnualMeanMaxTemp  0.02121855 -0.3730941  0.29744465
##                    logAnnualMeanMaxTemp
## Year                         0.02121855
## ayy                         -0.37309411
## logVap                       0.29744465
## logAnnualMeanMaxTemp         1.00000000
```

## 1.3 Local government areas (LGAs)

Table A1.3: Local government areas in Queensland

| LGA Code | LGA Name | Area of LGA (ha) | % forest in 1972 | Included in analysis | Li |
|---|---|---|---|---|---|
| 30250 | Aurukun | 734,721 | 63.3 | y | -1.85 |
| 30300 | Balonne | 3,110,618 | 27.7 | y | 1.56 |
| 30370 | Banana | 2,854,631 | 38.4 | y | 1.14 |
| 30410 | Barcaldine | 5,352,067 | 14.1 | y | 0.34 |
| 30450 | Barcoo | 6,182,501 | 0.5 | - | - |
| 30760 | Blackall Tambo | 3,038,906 | 28.7 | y | 0.53 |
| 30900 | Boulia | 6,095,581 | - | - | - |
| 31000 | Brisbane | 133,809 | 45.8 | y | 2 |
| 31750 | Bulloo | 7,376,280 | 0.5 | - | - |
| 31820 | Bundaberg | 643,564 | 53.7 | y | 1.99 |
| 31900 | Burdekin | 504,342 | 36.8 | - | - |
| 31950 | Burke | 4,003,921 | 4 | - | - |
| 32070 | Cairns | 411,511 | 70 | y | 0.97 |
| 32250 | Carpentaria | 6,412,490 | 4.8 | y | -1.97 |
| 32260 | Cassowary Coast | 468,499 | 62.2 | y | 1.99 |
| 32270 | Central Highlands | 5,983,489 | 41.9 | y | 0.68 |
| 32310 | Charters Towers | 6,837,355 | 48.5 | y | -0.89 |
| 32330 | Cherbourg | 3,160 | 65.9 | y | 1.63 |
| 32450 | Cloncurry | 4,798,313 | 0.6 | - | - |
| 32500 | Cook | ######## | 61.6 | y | -1.71 |
| 32600 | Croydon | 2,948,710 | 6.3 | y | -1.24 |
| 32750 | Diamantina | 9,466,685 | - | - | - |
| 32770 | Doomadgee | 183,516 | 2.5 | - | - |
| 33100 | Etheridge | 3,920,102 | 28.1 | - | - |
| 33200 | Flinders | 4,119,266 | 14.9 | y | -0.04 |
| 33220 | Fraser Coast | 710,250 | 64.7 | y | 1.99 |
| 33360 | Gladstone | 1,046,579 | 47.4 | y | 2 |
| 33430 | Gold Coast | 133,168 | 47 | y | 2 |
| 33610 | Goondiwindi | 1,925,549 | 24.4 | y | 2 |
| 33620 | Gympie | 688,454 | 51 | y | 1.99 |
| 33800 | Hinchinbrook | 280,135 | 52.9 | y | 0.72 |
| 33830 | Hope Vale | 110,475 | 58.5 | y | 0.81 |
| 33960 | Ipswich | 108,849 | 30 | y | 2 |
| 33980 | Isaac | 5,871,984 | 37.8 | y | 0.07 |
| 34420 | Kowanyama | 254,317 | 14.1 | - | - |
| 34570 | Lockhart River | 357,806 | 75.5 | - | - |
| 34580 | Lockyer Valley | 226,874 | 44.7 | y | 2 |
| 34590 | Logan | 95,810 | 51.2 | y | 2 |
| 34710 | Longreach | 4,057,171 | 3.6 | y | 0.97 |
| 34770 | Mackay | 760,120 | 56.7 | y | 0.6 |
| 34800 | McKinlay | 4,073,371 | 0.1 | - | - |
| 34830 | Mapoon | 54,799 | 61.3 | - | - |
| 34860 | Maranoa | 5,871,136 | 42.4 | y | 0.89 |
| 35010 | Moreton Bay | 203,330 | 58.5 | y | 2 |
| 35250 | Mornington | 124,421 | 36.7 | - | - |
| 35300 | Mount Isa | 4,318,804 | 0.9 | - | - |
| 35600 | Murweh | 4,069,849 | 51 | y | 0.59 |
| 35670 | Napranum | 199,791 | 62.4 | y | 0.64 |
| 35760 | North Burnett | 1,966,677 | 41.6 | y | 1.74 |
| 35780 | Northern Peninsula Area | 105,691 | 59.3 | - | - |
| 35790 | Palm Island | 7,063 | 43.3 | - | - |
| 35800 | Paroo | 4,761,641 | 17.3 | y | 0.69 |
| 36070 | Pormpuraaw | 442,884 | 27.7 | - | - |
| 36150 | Quilpie | 6,742,331 | 11.3 | y | -0.06 |
| 36250 | Redland | 53,625 | 51.1 | y | 2 |
| 36300 | Richmond | 2,658,022 | 3 | y | -2 |
| 36360 | Rockhampton | 1,831,174 | 44.6 | y | 1.47 |
| 36510 | Scenic Rim | 424,807 | 26.9 | y | 2 |
| 36580 | Somerset | 537,328 | 40.7 | y | 1.98 |
| 36630 | South Burnett | 838,170 | 28.8 | y | 2 |
| 36660 | Southern Downs | 711,172 | 31.4 | y | 2 |
| 36710 | Sunshine Coast | 312,071 | 56.2 | y | 2 |
| 36810 | Tablelands | 6,479,381 | 36.4 | y | -0.92 |
| 36910 | Toowoomba | 1,295,773 | 23 | y | 2 |
| 36950 | Torres | 88,271 | 38.9 | - | - |
| 36960 | Torres Strait Island | 48,924 | 29.7 | - | - |
| 37010 | Townsville | 372,738 | 60.7 | y | 1.85 |
| 37300 | Weipa | 1,082 | 11.2 | - | - |
| 37310 | Western Downs | 3,793,849 | 30.5 | y | 2 |
| 37340 | Whitsunday | 2,380,438 | 37.2 | y | -0.25 |
| 37400 | Winton | 5,381,439 | - | - | - |
| 37550 | Woorabinda | 39,028 | 50 | y | 1.69 |
| 37570 | Wujal Wujal | 1,118 | 80.8 | - | - |
| 37600 | Yarrabah | 15,884 | 78.7 | - | - |

## 1.4 Prior distributions and model variations

We specify the following Bayesian prior distributions:

$$b_0, b_{1k}, b_{2k}, b_3, a_k \sim N(\text{mean} = m_1, \text{var} = u) \text{ for all } k$$
$$\mathcal{T} \sim N(\text{mean} = m_2, \text{var} = 100)$$
$$\ell_\gamma \sim N(\text{mean} = \log 5.5, \text{var} = [\log 10]^2)$$
$$\nu^{-2}, \sigma_\tau^{-2}, \sigma_\gamma^{-2}, \sigma_k^2, \sigma_{k0}^{-2} \sim \text{Gamma}(1, 0.01) \text{ for all } k$$

where $m_1 = 16$ and $u = 100$ are the hyperparameter values for $b_0$, but $m_1 = 0$ and $u = 10$ for the other slope parameters; and $m_2$ corresponds to the model variations below. We chose our hyperparameter values above based on preliminary (exploratory) model fits, some of which appear in Appendix 1.5.

### 1.4.1 Timing of "bend" variation

To test whether the "bend" occurs around the year 2000 at the introduction of the original VMA, or around 2007 after the VMA amendments came into effect, we specify the hyperparameter values as $m_2 = 2000$ and $m_2 = 2007$, respectively.

### 1.4.2 Spatial adjacency weighting variation

To test if weighting spatial adjacency by land tenure similarity affects model fit, as compared to unweighted spatial adjacency, we specify the CAR structure in Level 2 of the modeling framework as:

$$\begin{cases} \beta_{10i} \mid \{\beta_{10j} : i \neq j\} \sim N\left(\text{mean} = \dfrac{\sum_{j \neq i} w_{ij} \beta_{10j}}{\sum_{j \neq i} w_{ij}}, \text{var} = \dfrac{\sigma_{10}^2}{\sum_{j \neq i} w_{ij}}\right) \\ \text{unweighted case: } w_{ij} = \mathbf{1}\{i, j \text{ share border}\} \text{ for all } i \neq j \\ \text{weighted case: } w_{ij} = \dfrac{\mathbf{1}\{i, j \text{ share border}\}}{|L_i - L_j| + 0.00001} \text{ for all } i \neq j \end{cases}$$

### 1.4.3 Other variations

Minor variations were considered and adopted for our final models if goodness-of-fit could be improved:

- alternative specification of prior distributions, e.g. bivariate joint prior for $\tau_i$ and $\log \gamma_i$
    - not adopted
- alternative hyperparameter values
    - not adopted
- alternative model parametrization in the WinBUGS implementation
    - not adopted
- reduced model complexity by taking $\log \gamma_i = \ell_\gamma$ for all $i$
    - adopted
- reduced model complexity by removing redundant driver variables
    - see Appendix 1.5

## 1.5 Justification for discarding model covariates

In addition to state-level scatterplots in Appendix 1.2, LGA-specific as well as year-specific scatterplots (e.g. year 2014 shown below) were additionally inspected.

From such inspection, land tenure was subsequently removed from consideration as a driver variable due to its redundancy with temperature throughout 1990 to 2014.

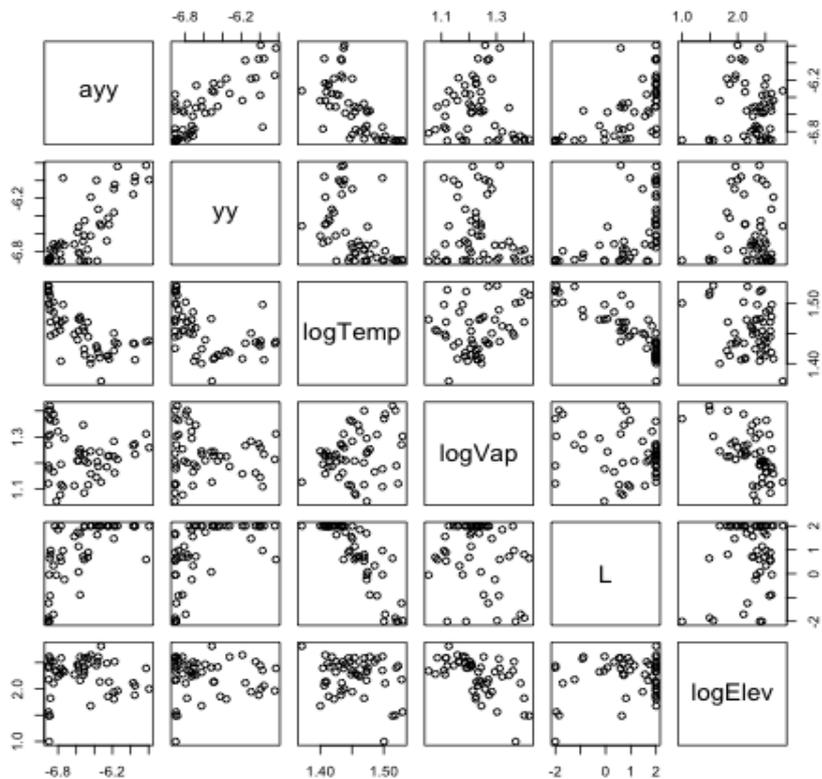

The remaining drivers were more formally assessed through preliminary model fits using WinBUGS or R-INLA [116]. Preliminary WinBUGS results suggested that Ag and Oil were highly redundant. Oil was kept for further assessment by R-INLA. Sample R-INLA results appear below. A smaller DIC value and effective number of parameters together suggest better goodness-of-fit. Thus, GDP was kept as the sole driver variable in our final models that are reported here and summarised in Appendix 1.6.

### 1.5.1 Unweighted CAR and iid year effect, multiple drivers and year:

```
summary(result495_21)
##
## Fixed effects:
##                mean     sd 0.025quant 0.5quant 0.975quant    mode kld
## (Intercept) 13.5073 8.3423    -2.9157  13.5077    29.9085 13.5091   0
## GDP         -0.0278 0.0203    -0.0677  -0.0278     0.0121 -0.0278   0
## oil         -0.0097 0.0351    -0.0788  -0.0097     0.0594 -0.0097   0
## temp        -0.0635 0.0629    -0.1863  -0.0637     0.0605 -0.0641   0
## vap          0.0813 0.0569    -0.0304   0.0812     0.1932  0.0810   0
## elev        -0.0555 0.0541    -0.1617  -0.0556     0.0513 -0.0558   0
## Year        -0.0099 0.0042    -0.0181  -0.0099    -0.0017 -0.0099   0

## Deviance Information Criterion (DIC) ...: -4.243
## Effective number of parameters .........: 73.84
```

### 1.5.2 Unweighted CAR and iid year effect, single driver and year:

```
summary(result495_2)
##
## Fixed effects:
##                mean     sd 0.025quant 0.5quant 0.975quant    mode kld
## (Intercept) 16.5864 3.9808     8.7707  16.5862    24.3954 16.5864   0
## GDP         -0.0268 0.0155    -0.0572  -0.0268     0.0036 -0.0268   0
## Year        -0.0115 0.0020    -0.0154  -0.0115    -0.0076 -0.0115   0

## Deviance Information Criterion (DIC) ...: -9.363
## Effective number of parameters .........: 71.45
```

## 1.6 Goodness-of-fit values for final models

Table A2.3. Model fits[8] for identically parametrized models: GDP and year are only covariates, with LGA-specific incoming and outgoing slopes, LGA-specific $\tau_i$, but common γ. Note that small deviance together with small $p_V$ imply good fit. As all four models involve the same number of parameters, the deviance can be used to compare all four against each other. In contrast, we do not use $p_V$ to compare between a Simple and a Weighted model, because weighting appears to cause a drastic increase in the instability[9] of the WinBUGS numerical approximation algorithms.

| Model variation | hyperparameter $m_2$ | Spatial weight $w_{ij}$ | Posterior median deviance | $p_V$ estimate[10] |
|---|---|---|---|---|
| Simple CAR | 2000 | $\mathbf{1}\{i,j \text{ share border}\}$ | -929.5 | 369616.0 |
| Simple CAR | 2007 | | -926.45 | 350471.5 |
| Weighted CAR | 2000 | $\dfrac{\mathbf{1}\{i,j \text{ share border}\}}{\mid L_i - L_j \mid +0.00001}$ | -1487.0 | 556042.4 |
| Weighted CAR | 2007 | | -1516.0 | 535166.6 |

---

[8] The results reported here are based on locally converged models, which may differ from the global best fit. Having a mere 50 LGAs in the dataset was the likely reason for WinBUGS to exhibit (a) unstable estimates of $p_V$ and (b) difficulty in locating the global best fit; indeed multiple local minima are inherent in the bent-cable model deviance [63,64].

[9] As suggested by the unusually large $p_V$ estimates.

[10] Computed by dividing the posterior variance of the deviance by two [101]

## 1.7 Influence of land tenure similarity weighting on spatial estimates

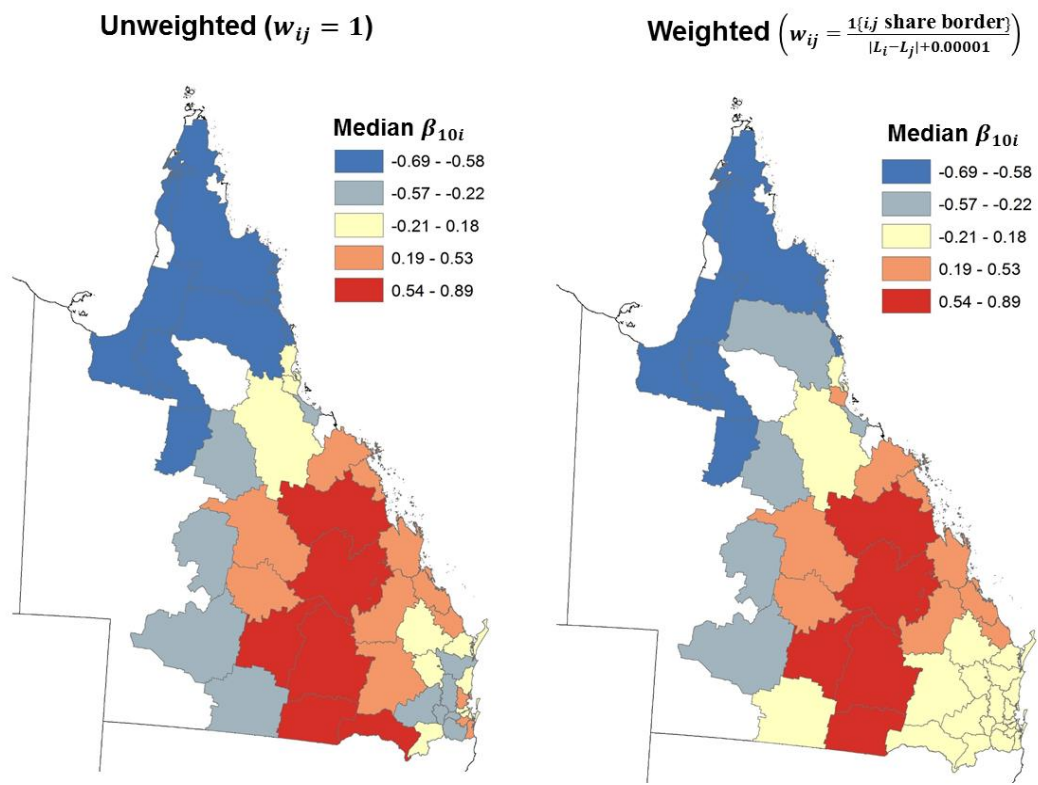

Figure A1.2. Estimates (posterior medians) of spatial random effects $\beta_{10i}$ for unweighted model (left, median deviance ~ -930) and weighted model (right, median deviance ~ -1500). Smaller deviance indicates better goodness-of-fit.

## 1.8 LGA specific model outputs

Figure A1.3: LGA-level time series $y_{it}$ ($n = 50$) after detrending (black), with LGA-level bent-cable estimate (red) and fitted population-level bent cable (green).

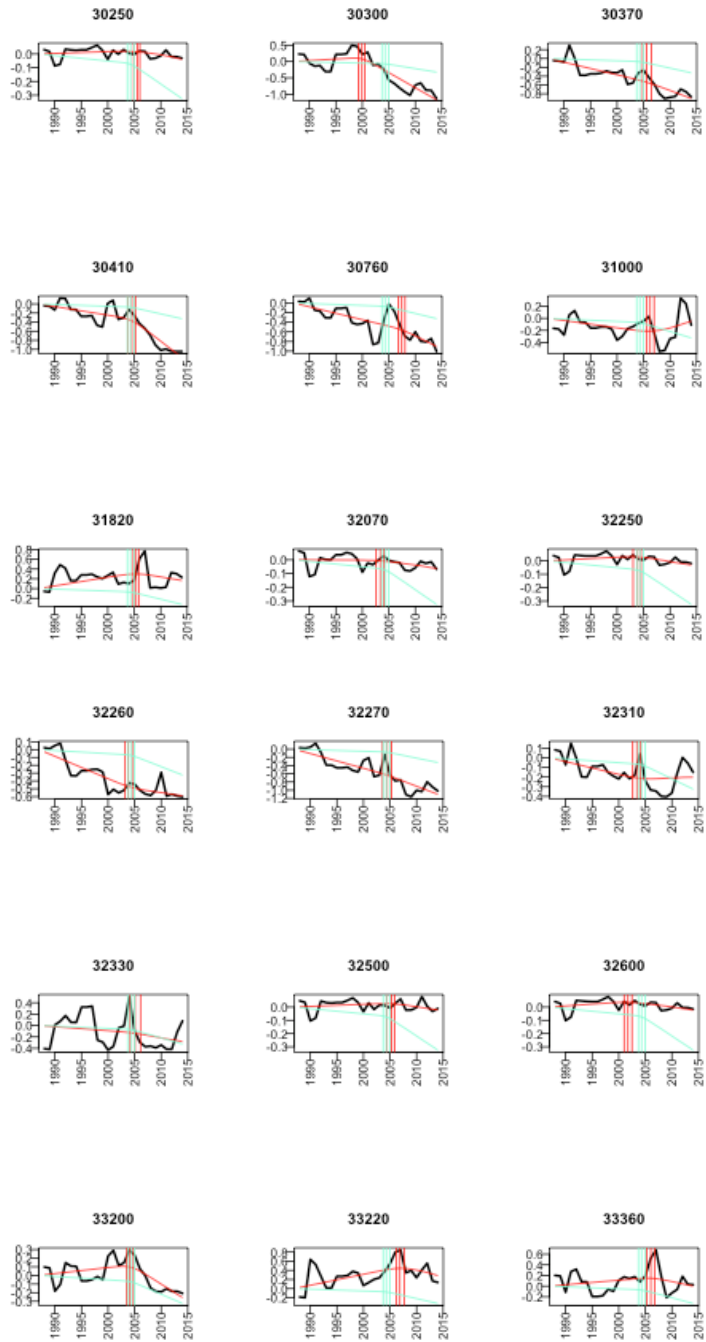

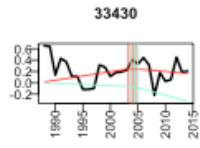 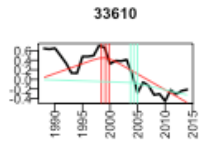 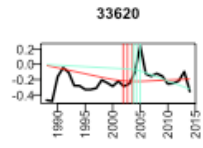
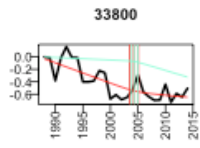 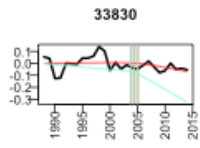 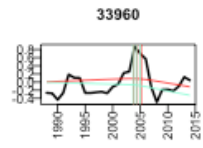
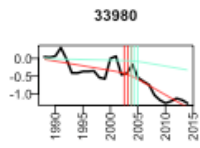 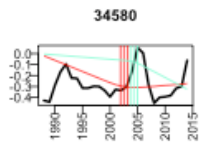 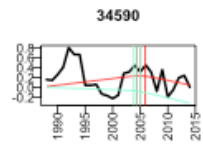

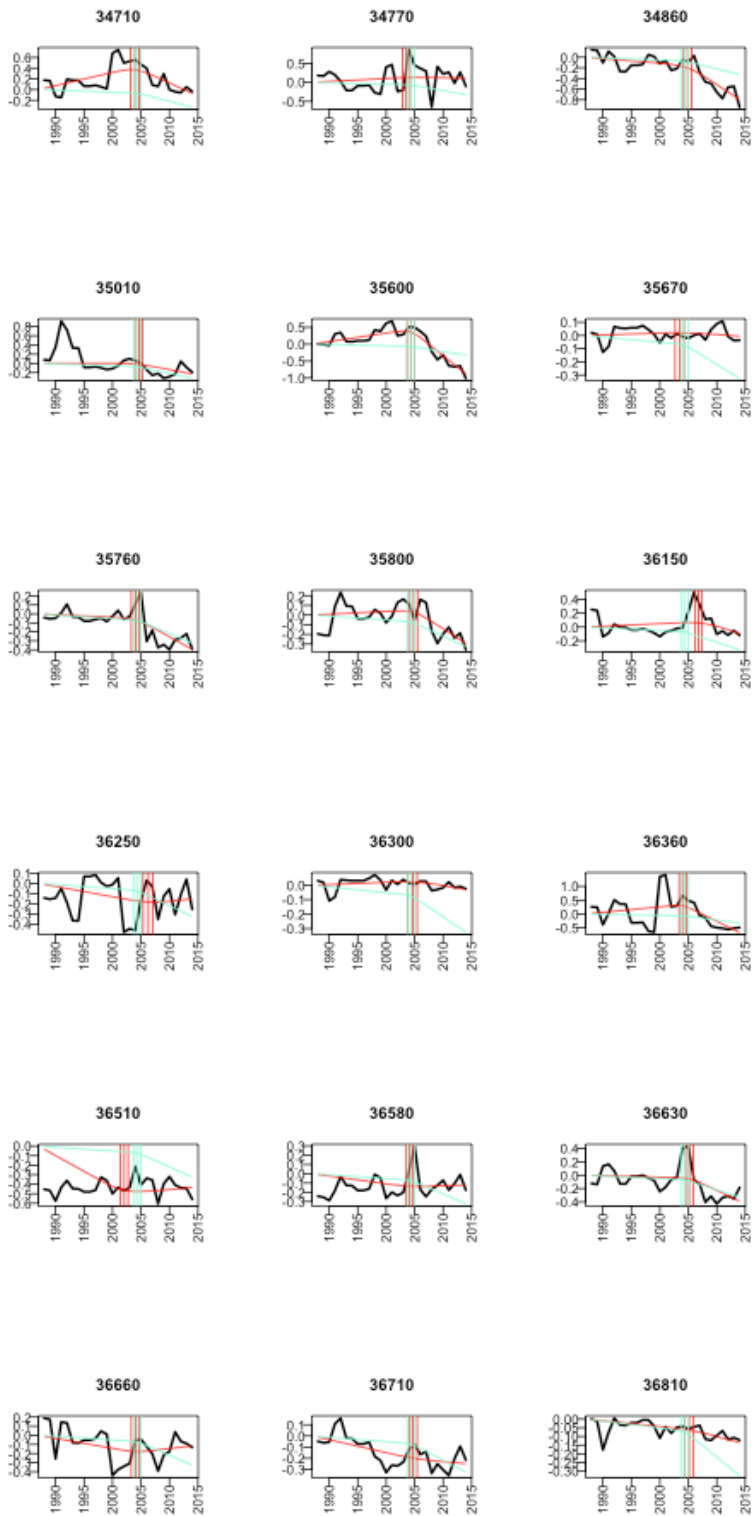

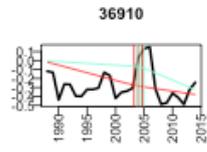 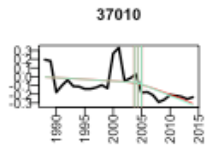 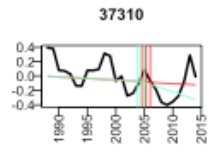

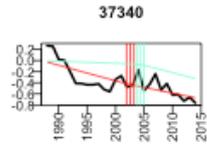 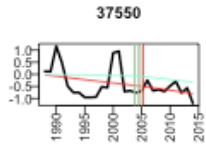